\documentclass[%
11pt,
superscriptaddress,
notitlepage,
nofootinbib,
amsmath,amssymb,
aps,
prd,
]{revtex4-1}

\usepackage{graphicx}
\usepackage{caption}
\usepackage{subcaption}
\graphicspath{{.}{./data/Figs/}}
\usepackage{float}% to force figure placement 
\usepackage{bm}% bold math
\usepackage[hidelinks]{hyperref}
\usepackage{xcolor}
\usepackage{upgreek}
\usepackage[utf8]{inputenc}
\usepackage[T1]{fontenc}
\usepackage{slashed}

\usepackage{fullpage}
\usepackage{enumerate}

\usepackage{mathalfa,amssymb,amsthm}
\usepackage{amsmath}

\begin{document}

%\preprint{APS/123-QED}

\title{Nonlinear studies of binary black hole mergers 
in Einstein-scalar-Gauss-Bonnet gravity}
\author{Maxence Corman}
\email{mcorman@perimeterinstitute.ca}
\affiliation{%
Perimeter Institute for Theoretical Physics, Waterloo, Ontario N2L 2Y5, Canada.
}%
\affiliation{%
Department of Physics and Astronomy, University of Waterloo, Waterloo, Ontario N2L 3G1,
Canada
}%
\author{Justin L. Ripley}
\email{ripley@illinois.edu}
\affiliation{%
Illinois Center for Advanced Studies of the Universe and Department of Physics,
University of Illinois at Urbana-Champaign, Urbana, Illinois 61801, USA
}%
\affiliation{%
DAMTP,
Centre for Mathematical Sciences,
University of Cambridge,
Wilberforce Road, Cambridge CB3 0WA, United Kingdom.
}%
\author{William E. East}
\email{weast@perimeterinstitute.ca}
\affiliation{%
Perimeter Institute for Theoretical Physics, Waterloo, Ontario N2L 2Y5, Canada.
}%

%-----------------------------------------------------------------------------

\date{\today}% It is always \today, today,

\begin{abstract}
   We study the nonlinear dynamics of binary black hole systems with scalar
   charge by numerically evolving the full equations of motion for
   shift-symmetric Einstein scalar Gauss-Bonnet gravity.  We consider
   quasi-circular binaries with different mass-ratios, varying the
   Gauss-Bonnet coupling and quantifying its impact on the emitted scalar and
   gravitational waves.  We compare our numerical results to post-Newtonian
   calculations of the radiation emitted during the inspiral.  
   We demonstrate the accuracy of the leading-order terms in post-Newtonian
   theory in modeling the amplitude of the scalar waveform, but find that, 
   at least for the
   last few orbits before merger, the currently available post-Newtonian
   theory is not sufficient to model the dephasing of the gravitational wave
   signal in this theory. We further find that there is non-negligible nonlinear
   enhancement in the scalar field at merger, but that the effect on
   the peak gravitational wave emission is small. 
\end{abstract}

\maketitle

%=============================================================================
%\allowdisplaybreaks
%\tableofcontents
%=============================================================================
\section{Introduction}
\label{sec:introduction}
%=============================================================================
In recent years, 
the gravitational waves (GW) observed from the inspiral, merger, and ringdown
of black hole binaries have greatly constrained the landscape
of potential deviations from General Relativity (GR)
   \cite{TheLIGOScientific:2016src,
   Yunes:2016jcc,
   Baker:2017hug,
   Abbott:2018lct, 
   Isi:2019aib, 
   Abbott:2020jks, 
   Isi:2020tac, 
   Psaltis:2020lvx, Volkel:2020xlc, Kocherlakota:2021dcv,
   Okounkova:2021xjv}.
However, in order to seek physics beyond GR, or to place the most
stringent constraints on deformations of GR, one needs 
accurate predictions for specific modified gravity theories, 
in particular in the strong field and dynamical regime
\cite{Yunes:2013dva,Berti:2015itd,Berti:2018cxi,Berti:2018vdi}.
This has been a major theoretical and technical 
challenge for many theories of interest
\cite{Hirschmann:2017psw,Okounkova:2017yby,Okounkova:2019dfo,Okounkova:2020rqw,
Cayuso:2017iqc,Cayuso:2020lca,
East:2020hgw,Ripley:2022cdh,Franchini:2022ukz,Bezares:2021dma}. 
As a result, most tests of GR performed so far are model-independent
or null tests, more commonly classified as
consistency and parametrized tests \cite{Abbott:2018lct,
LIGOScientific:2019fpa,
LIGOScientific:2020tif,LIGOScientific:2021sio,Ghosh:2022xhn}.
Parametrized tests introduce deviations from GR to the gravitational
waveform in a theory-agnostic way,
and use the data to constrain the beyond GR parameters.
Most current approaches, however, usually only constrain
the deviations by considering one specific modification
at a time and thus the interpretation of these constraints
remains limited.

An interesting class of theories to test against GR is 
Einstein-scalar-Gauss-Bonnet (ESGB) gravity,
which introduces modifications to GR at small curvature length
scales.
Variants of ESGB gravity allow for scalar-charged black holes
\cite{Kanti:1997br,Yunes:2011we,Sotiriou:2013qea,Sotiriou:2014pfa},
and hence can differ qualitatively from GR in the strong field regime,
while still passing weak field tests.
Because of this, much recent work has gone into modeling 
compact object mergers
in ESGB gravity in both post-Newtonian (PN) theory 
\cite{Yagi:2011xp,Sennett:2016klh,Shiralilou:2020gah,Shiralilou:2021mfl}
and numerical relativity
\cite{Witek:2018dmd,Silva:2020omi,Okounkova:2020rqw,
East:2020hgw,East:2020hgw,Ripley:2022cdh,East:2022rqi}.
In Ref.~\cite{East:2020hgw}, two of us introduced a computational
methodology to solve the equations of motion for binary black
hole system in ESGB gravity without
approximation (beyond that of numerical truncation error), by making use
of the modified generalized harmonic (MGH) formulation
\cite{Kovacs:2020pns,Kovacs:2020ywu}.\footnote{These methods were recently extended to a modified version of the CCZ4 formulation
in Ref.~\cite{AresteSalo:2022hua}.}
Here we follow up on that work, 
and study 
the dynamics of the last stages of the inspiral phase of
quasi-circular, non-spinning black holes in shift-symmetric ESGB (sGB) gravity,
and investigate the accuracy of PN approximations 
\cite{Yagi:2011xp,Sennett:2016klh,Shiralilou:2020gah,Shiralilou:2021mfl}.

In general, the equations of motion for ESGB gravity can only be stably
evolved in time for weakly-coupled solutions 
\cite{Ripley:2019irj,Kovacs:2020pns,Kovacs:2020ywu,East:2020hgw,Ripley:2022cdh}.
Weak coupling roughly means that the Gauss-Bonnet corrections
to the spacetime geometry remain sufficiently 
small compared to the smallest curvature length scale in the solution.
A binary black hole system in ESGB gravity can evolve from an initially
weakly coupled state to a strongly coupled state, as the black holes
become closer and eventually merge \cite{Julie:2019sab,Julie:2022huo}.
We find that in a significant portion of the parameter space,
our evolution breaks down as the black holes
become closer, although approaching this limit does not appear
to be preceded by dramatically different spacetime or scalar
field dynamics compared to the weakly-coupled regime.
Maintaining a weakly-coupled solution exterior to the black hole
horizons through merger remains a major challenge in the numerical
evolution of binary black holes in numerical relativity.
While better addressing this issue remains an important issue
for future work, for many cases here we focus on the properties
of the late inspiral phase of binary evolution.
Even when restricting to the inspiral phase, we show
that the deviations from GR are significant in terms
of the imprint on the resulting gravitational waves.
One of our main results is that leading order
PN approximations are not sufficient
to model the gravitational signal in the late stages of the inspiral.
For the cases we were able to evolve through merger,
we find that the effects of ESGB gravity 
show up primarily in an nonlinear enhancement of the scalar field
at merger, 
and in the dephasing of the gravitational waves,
while the effect on the peak amplitude of the gravitational wave signal
is small.
This work also demonstrates the efficacy of the numerical relativity 
techniques utilized here---which 
should be applicable to any scalar-tensor theory with second order
equations of motion---to quantify the impact on the gravitational wave
signal of modified gravity in regimes where other approximations break down. 

The remainder of the paper is as follows. In Sec.~\ref{sec:eom_derivation}, we
review shift-symmetric ESGB gravity. In Sec.~\ref{sec:numerical_methods}, we
describe our numerical methods for evolving this theory and analyzing the
results.  Results from our study of quasi-circular binary black holes in sGB
are presented in Sec.~\ref{sec:results}.  We discuss these results and conclude
in Sec.~\ref{sec:conclusion}. We discuss the accuracy of our simulations in
Appendix~\ref{sec:error_analysis}, collect PN results in sGB in
Sec.~\ref{sec:pn_theory}, outline our initial-data set-up in
Appendix~\ref{sec:puncture_id}, and review the accuracy of the perturbative
approach to solving the equations of motion
in Appendix~\ref{sec:perturbative_method}.  We use geometric units:
$G=c=1$, a metric sign convention of $-+++$, lower case Latin letters to index
spacetime indices, and lower case Greek letters to index spatial indices.  
The Riemann tensor is $R^a{}_{bcd}=\partial_c\Gamma^a_{db}-\cdots$.
%=============================================================================
\section{Shift-symmetric ESGB gravity}
\label{sec:eom_derivation}
We briefly review shift-symmetric ESGB (sGB) gravity. The action is:
\begin{align}
\label{eq:leading_order_in_derivatives_action}
    S
    =&
    \frac{1}{16\pi}\int d^4x\sqrt{-g}
   \left(
        R
    -   \left(\nabla\phi\right)^2
    +	2\lambda\phi\mathcal{G}
    \right)
    ,
\end{align}
    where $\mathcal{G}$ is the Gauss-Bonnet scalar:
\begin{align}
    \mathcal{G}
    \equiv
    R^2 - 4R_{ab}R^{ab} + R_{abcd}R^{abcd}
    .
\end{align}
Here, $\lambda$ is a constant coupling parameter that, in
geometric units, has dimensions of length squared.
As the Gauss-Bonnet scalar $\mathcal{G}$ is a total derivative
in four dimensions, we see that the action of 
sGB gravity is preserved
up to total derivatives under constant shifts in the scalar field:
$\phi\to\phi+\textrm{constant}$.
Schwarzschild and Kerr black holes are not stationary solutions 
in this theory: if one begins with such vacuum initial data, the black holes will dynamically develop stable scalar clouds (hair).
The end state then is a scalar-charged black hole,
so long as the coupling normalized by the black hole mass $m$,
$\lambda/m^2$, is sufficiently small
\cite{Sotiriou:2013qea,Sotiriou:2014pfa,Ripley:2019aqj,East:2020hgw}.
In particular, 
regularity of black hole solutions and hyperbolicity of the theory sets 
$\lambda/m^2\lesssim 0.23$ for non-spinning black holes, 
\cite{Sotiriou:2014pfa,Ripley:2019aqj}.
In contrast to stars, where the scalar field
around them falls of more rapidly than $1/r$,
black holes have a scalar charge, and thus
black hole binaries emit scalar radiation, 
which increases the speed at which the binary inspirals and merges 
\cite{Yagi:2015oca,Yagi:2011xp}.
The most stringent observational bounds on the theory
come from unequal mass, or black hole-star binaries,
as those emit scalar dipole radiation, which leads to a more rapid
dephasing of the gravitational waveform than would be observed in GR.
In PN theory, the scalar dipole radiation enters as a $-1$PN effect and
can dominate over gravitational radiation at sufficiently wide separations (low frequencies).
In this study, we will focus on late inspiral, where the gravitational waves are strongest and
the scalar radiation is subdominant (the quadrupolar driven inspiral regime).
Another feature of these solutions is that the scalar charge 
is inversely proportional to the square of the smallest mass black hole in the system.
This suggests that the best way to probe EsGB gravity is by observing the
smallest compact objects.  We therefore expect stronger constraints on the
theory will come from observing the merger of stellar mass black holes with
ground-based detectors, as opposed to observations of supermassive black hole
mergers with LISA (although long-duration observations of extreme mass-ratio
insipirals with LISA may provide meaningful
constraints~\cite{Chamberlain:2017fjl}).
Restoring dimensions, comparisons of gravitational wave observations from
the LIGO-Virgo-KAGRA catalogue to PN results place constraints of 
$\sqrt{\lambda}\lesssim 2.5$ km,
see Refs.~\cite{Perkins:2021mhb,Lyu:2022gdr}. 

%=============================================================================
\section{Methods}
\label{sec:numerical_methods}
%------------------------------------------------------------------------------
\subsection{Evolution equations and code overview}
\label{sec:evolution_methods}

The covariant equations of motion for sGB gravity are
\begin{align}
\label{eq:eom_esgb_scalar}
   \Box\phi
   +  
   \lambda\mathcal{G}
   &=
   0
   ,\\
\label{eq:eom_edgb_tensor}
   R_{ab}
   -  
   \frac{1}{2}g_{ab}R
   -  
   \nabla_a\phi\nabla_b\phi
   +  
   \frac{1}{2}\left(\nabla\phi\right)^2g_{ab}
   +  
   2\lambda
   \delta^{efcd}_{ijg(a}g_{b)d}R^{ij}{}_{ef}
   \nabla^g\nabla_c\phi
   &=
   0
   ,
\end{align}
where $\delta^{abcd}_{efgh}$ is the generalized Kronecker delta tensor.
We numerically evolve the full sGB equations of motion
using the MGH 
formulation~\cite{Kovacs:2020pns,Kovacs:2020ywu}.
We use similar choices for the gauge and numerical parameters as in
Ref.~\cite{East:2020hgw}.
We worked with box-in-box adaptive mesh refinement as provided by
the PAMR library \cite{PAMR_online}. We typically worked with eight 
levels of mesh refinement in our simulations, unless otherwise noted.
We provide details on numerical resolution and convergence in
Appendix~\ref{sec:error_analysis}.
%------------------------------------------------------------------------------
\subsection{Puncture binary black hole initial data\label{sec:initial_data}}
On our initial time slice, we must satisfy the generalizations
of the Hamiltonian and momentum constraint equations to sGB. 
Here, we do not implement
a method to solve the equations for general $\phi$, but instead
consider initial data for which $\phi=\partial_t\phi=0$.
With this choice of $\phi$,
the constraint equations of sGB gravity reduce to those of vacuum GR 
\cite{East:2020hgw,Ripley:2022cdh}.
Even though $\phi=\partial_t \phi=$0 on the initial time slice,
scalar field clouds subsequently form on a timescale that is short compared with the 
orbital binary timescale (within $\sim100M_0$). 
We construct quasi-circular binary black hole initial data
via the black hole puncture method
\cite{Brandt:1997tf}, using the \texttt{TwoPunctures} code 
\cite{Ansorg:2004ds,Paschalidis:2013oya}\footnote{The 
particular version of the code we use can be accessed at \cite{code_repo_tp}.}.

For puncture binary black hole
initial data, we need to specify the initial black hole positions,
and their approximate initial masses $m_{1,2}$
(with the convention that $m_1\leq m_2$), linear momenta
$P^{\gamma}_{1,2}$, and spins $S^{\gamma}_{1,2}$ (which we set to zero in this study).
Given $m_{1,2}$ and the initial puncture (black hole) locations, 
we use the dynamics for a circular binary to 2PN order to determine
the tangential components to $P^{\gamma}_{1,2}$, and the 
2.5PN radiation reaction term to determine the initial radial
component of $P^{\gamma}_{1,2}$~\cite{vasilis_notebook}.
We review our initial data setup in more detail in
Appendix~\ref{sec:puncture_id}.

For the first $t=50M_0$ (where $M_0\equiv m_1+m_2$) of evolution,
we evolve the black holes purely in GR. We found this allowed for
the junk radiation from the puncture initial data to disperse
away from the black holes. After that initial evolution time,
we turn on the Gauss-Bonnet coupling $\lambda$ to a
non-zero value. The constraints are satisfied in this procedure,
as we can think of our initial data as starting at
$t=50M_0$ instead, with $\phi=\partial_t\phi=0$ and a metric field that
satisfies the constraints such that the initial data satisfies
the constraint equations for sGB gravity
\cite{East:2020hgw,Ripley:2022cdh}.
While we use quasi-circular initial data based on PN approximations
for the initial orbital velocities 
from GR, we found that the scalarization process does not appreciably
impact the eccentricity of our runs, and instead
the eccentricity of our runs is dominated by the truncation
error of the simulations.  
For more discussion, see Appendix~\ref{sec:error_analysis}.

%------------------------------------------------------------------------------
\subsection{Diagnostic Quantities\label{sec:diagnostics}}

We use many of the same diagnostics as in 
Ref.~\cite{East:2020hgw}, which we briefly review here.
We measure the scalar and gravitational
radiation by extracting the scalar field $\phi$ and Newman-Penrose
scalar $\Psi_4$ on finite-radius coordinate spheres.
Due to the coupling between the scalar field and metric through
the Gauss-Bonnet coupling,
in general scalar and gravitational radiation will couple together
through the term $\delta \times Riemm \times \nabla\nabla\phi$.
For asymptotically flat spacetimes that have an asymptotically
flat future null infinity
(that is spacetimes for which the peeling theorem holds, so the
Weyl scalar fall of sufficiently fast \cite{NP_formalism_paper}), 
this coupling falls off as $1/r^4$ as $r\to\infty$. 
For those spacetimes, in the wave zone, we can treat the
gravitational and scalar radiation as two uncoupled quantities
(for related discussions, see
\cite{Tattersall:2018map,East:2020hgw}).
We discuss how we estimate the finite-radius extraction error of our waveforms
in Appendix~\ref{sec:error_analysis}.

We decompose
$\Psi_4$ and $\phi$ into their spin-weighted spherical harmonic components
\begin{subequations}
\begin{align}
   \label{eq:psi4_decomposition}
   \Psi_{4,\ell m}(t,r)
   &\equiv
   \int_{\mathbb{S}_2} {}_{-2}\bar{Y}_{\ell m}\left(\vartheta,\varphi\right)
   \Psi_4\left(t,r,\vartheta,\varphi\right)
   ,\\
   \label{eq:phi_decomposition}
   \phi_{\ell m}(t,r)
   &\equiv
   \int_{\mathbb{S}_2} {}_{0}\bar{Y}_{\ell m}\left(\vartheta,\varphi\right)
   \phi\left(t,r,\vartheta,\varphi\right)
   .
\end{align} 
\end{subequations}
The gravitational wave luminosity is 
\begin{align}
\label{eq:radiated_gw_power}
   P_{\rm GW}(t)
   =
   \lim_{r\to\infty}\frac{r^2}{16\pi}
   \int_{\mathbb{S}_2}
   \left|\int_{-\infty}^t\Psi_4\right|^2
   .
\end{align} 
The scalar wave luminosity is $P_{\rm SF}$
\begin{align}
\label{eq:radiated_scalar_power}
   P_{\rm SF}
   \equiv
   -
   \lim_{r\to\infty}
   r^2
   \int_{\mathbb{S}_2} 
   N t^a\left(T^{\rm SF}\right)^b_adA_b
   ,
\end{align}
where $N=1/\sqrt{-g^{tt}}$ is the lapse and $t^a$ is the asymptotic
timelike Killing vector, the integral is over a sphere, and
\begin{align}
   T^{\rm SF}_{ab}
   \equiv
   \frac{1}{8\pi}\left(
      \nabla_a\phi\nabla_b\phi
      -
      \frac{1}{2}g_{ab}\nabla_c\phi\nabla^c\phi
   \right)
   .
\end{align}
We assume the scalar radiation is outgoing, 
so that Eq.~\eqref{eq:radiated_scalar_power} reduces to 
\begin{align}
\label{eq:radiated_scalar_power_2}
   P_{\rm SF}(t)
   &=
   \lim_{r\to\infty}\frac{r^2}{8\pi}
   \int_{\mathbb{S}_2}
    \left(\partial_t{\phi}\right)^2
   .
\end{align}

To compare our numerical waveforms,
we must estimate the orbital frequency of the binary $\Omega$.
We do so using the approximate relation 
\cite{PhysRev.131.435,Berti:2007fi,Maggiore:2007ulw}
\begin{align}
   \label{eq:orbital_frequency}
   \Omega
   \approx
   \frac{1}{2}\frac{d\Phi_{22} (t)}{dt}
   ,
\end{align}
where $\Phi_{22}/2$ is the definition of orbital phase 
computed from half the complex phase of $\Psi_{4,22}$.
We track the apparent horizons (AHs) associated with the black holes,
and measure their areas and associated angular momentum $J_{\rm BH}$. 
From this, we compute the black hole mass $m_{\rm BH}$
via the Christodoulou formula \cite{Christodoulou:1970wf}
\begin{align}
   m_{\rm BH}
   \equiv
	\sqrt{M_A^2 + \frac{J_{AH}^2}{4M_A^2}}
   ,
\end{align}
where $M_A\equiv \sqrt{\mathcal{A}/(16\pi)}$ is the areal mass.
We note that while the areal mass always increases
in vacuum GR \cite{hawking_area_original}, 
it can decrease in sGB gravity as the theory
can violate the Null Convergence Condition
(which is $R_{ab}k^ak^b\geq0$ for all null $k^a$) 
\cite{Ripley:2019irj,Ripley:2019aqj}.
In our simulations, $J_{\rm AH}\approx 0$ to numerical precision for the constituents of the binary black hole.
We measure the average value of the
scalar field on the black hole apparent horizons
\begin{align}
   \left<\phi\right>_{AH}
   \equiv
	\frac{1}{\mathcal{A}}\int_{AH}\phi
   .
\end{align}

%------------------------------------------------------------------------------
\subsection{Cases considered\label{sec:cases}}
We focus on quasi-circular black hole binaries with no spin.
We classify our runs by two dimensionless numbers:
their mass ratio $q$ and by the relative
Gauss-Bonnet scalar coupling strength $\zeta_1$ 
(compare to Refs.~\cite{Yagi:2011xp,Perkins:2021mhb,Lyu:2022gdr}): 
\begin{align}
   q
   \equiv
   \frac{m_1}{m_2}
   \leq 1
   ,\qquad
   \zeta_1
   \equiv
   \frac{\lambda}{m_1^2}
   .
\end{align}
As $m_1$ is the smaller black hole mass, it roughly quantifies the smallest
curvature scale in our simulations. We consider the mass ratios  $q=1$, $2/3$, and $1/2$,
with an initial separation of 10$M_0$, approximately 8 orbits before merger in GR.
For the equal mass ratios, we consider ESGB coupling parameters 
$\zeta_1= 0$, 0.01, 0.05, and $0.1$; while for the mass ratios $q=2/3$ and $q=1/2$,
we consider smaller values of $\zeta_1 = 0$, 0.025, 0.05, and 0.075; and 
$\zeta_1 = 0$, 0.05, and 0.075, respectively.
When comparing waveforms ($\Psi_{4,\ell m}$ or $\phi_{\ell m}$) with different values of the coupling,
we compute the time $t_{\rm align}$ at which the gravitational wave frequency is $0.01 M_0$,
and apply this as a time offset. This alleviates the effect of any dephasing or shift in frequency due to the
scalarization process.
We then rotate
the waveforms by a constant, complex phase so that their initial
phases align.
For comparisons with other works, our coupling $\lambda$ corresponds to 
$\alpha_{\rm GB} \equiv \lambda/\sqrt{8\pi}$ used in, e.g.
\cite{Perkins:2021mhb,Lyu:2022gdr}.\footnote{
 However, several other studies 
 (e.g.~\cite{Witek:2018dmd,Blazquez-Salcedo:2016enn,Pierini:2021jxd,Pierini:2022eim})
 take conventions leading to a value of $\alpha_{\rm GB}$
 that is $16\sqrt{\pi}\times$ times larger.
}
Restoring physical units, we have
\begin{align}
   \sqrt{\alpha_{\mathrm GB}} 
   \approx
   3.97 \ 
   \mathrm{km}
      \left(\frac{\sqrt{\lambda}}{m_1}\right)\left(\frac{m_1}{6\ M_{\odot}}\right)
\end{align}
where $6\ M_{\odot}$ is approximately the value of smallest black hole 
observed in the LIGO-Virgo-Kagra third observing run
\cite{LIGOScientific:2021djp}. For reference, Ref.~\cite{Lyu:2022gdr} sets a constraint 
of $\sqrt{\alpha_{\rm GB}} \lesssim 1.2$ km by comparing gravitational wave
observations of black hole-neutron star binaries to PN results of ESGB.
In comparison, the largest coupling we consider in our simulations 
(our equal-mass $\zeta=0.1$ run) corresponds to
$\sqrt{\alpha_{\rm GB}} \sim 1.25$ km for a $6\ M_{\odot}$ black hole,
which is roughly within observational bounds.

%------------------------------------------------------------------------------
\subsection{Challenges in modeling the
merger phase of black hole evolution\label{sec:merger}}
As we discuss in Sec.~\ref{sec:results}, 
we are unable to evolve the binaries through merger for many of our simulations.
For some of our runs, we turned off the scalar Gauss-Bonnet coupling
inside a compact ellipsoidal region centered at the black hole binaries
center of mass at a finite time before merger. 
This allowed us to evolve through merger, and extract gravitational
and scalar radiation from the inspiral up until the causal future
of the excised region intersected where we measured the radiation 
(typically at $r/M_0=90$).

For one case, namely $q=1$ and $\zeta_1=0.05$, we only turn off the
Gauss-Bonnet coupling slightly before finding a common apparent horizon, and
only in a localized region that is encompassed by the final black hole. We have
verified that varying the size of this region has no appreciable impact on the
resulting radiation, and so we include the full results from this case, though
a careful tracking of the propagation of information along
characteristics would be needed to more rigorously justify this.

We believe the main difficulty with evolving through merger in our
simulations may be elliptic regions that form around merger.
These regions may possibly be hidden behind the final event horizon,
and so could possibly be excised from the computational domain
if an apparent horizon is located quickly enough.
Higher resolution runs, 
with excision surfaces that lie closer to the
apparent horizons of the inspiraling black holes
or a different choice of the auxiliary metrics in the modified generalized
harmonic formulation, may allow for
the successful merger of black holes in sGB gravity with unequal
mass ratios.
We leave a further investigation of this to future work.

%=============================================================================
\section{Results\label{sec:results}}
We present results for binary black holes with several mass ratios, beginning
roughly eight orbits before merger, focusing on how the orbital dynamics and
radiation changes as a function of the sGB coupling.
We compare both the scalar radiation, and the modified gravity induced dephasing
of the orbit and gravitational wave signal to the PN prediction. 

%=============================================================================
\subsection{Scalar radiation and dynamics\label{sec:scalar_results}}
\begin{figure*}
\centering
   \begin{subfigure}{0.45\textwidth}
      \includegraphics[width=\columnwidth,draft=false]{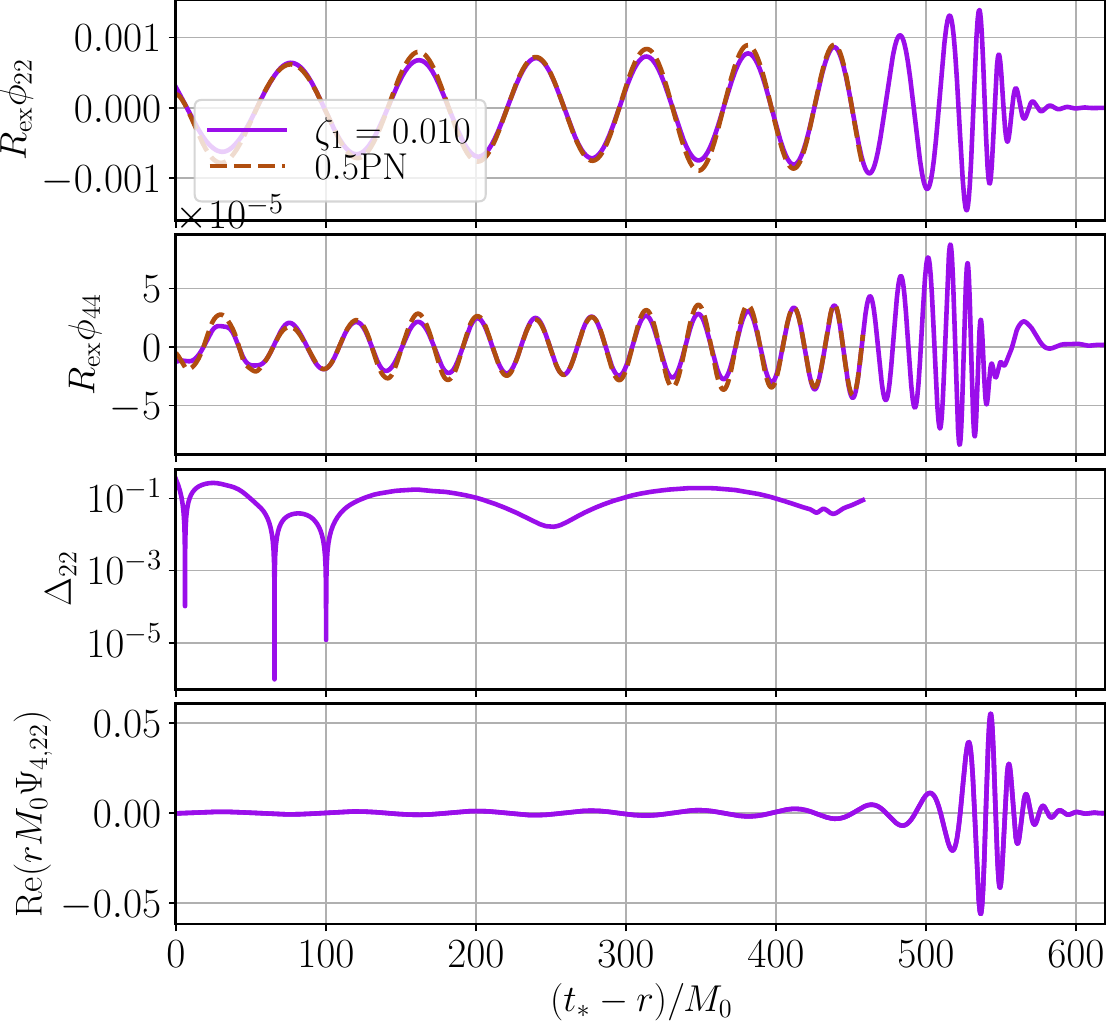}
      \caption{$q=1$}
   \end{subfigure}
   \begin{subfigure}{0.45\textwidth}
      \includegraphics[width=\columnwidth,draft=false]{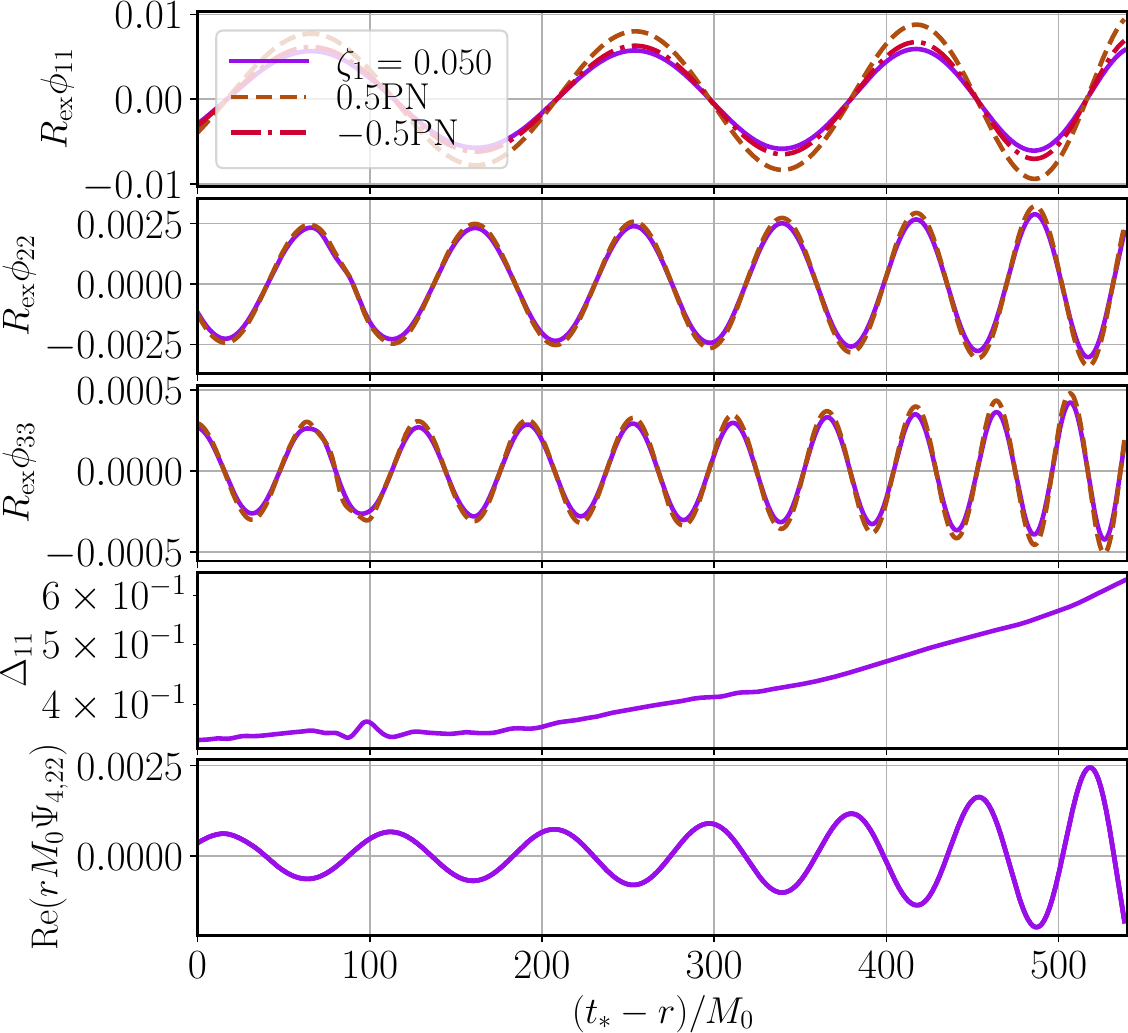}
      \caption{$q=2/3$}
   \end{subfigure}
   \begin{subfigure}{0.45\textwidth}
      \includegraphics[width=\columnwidth,draft=false]{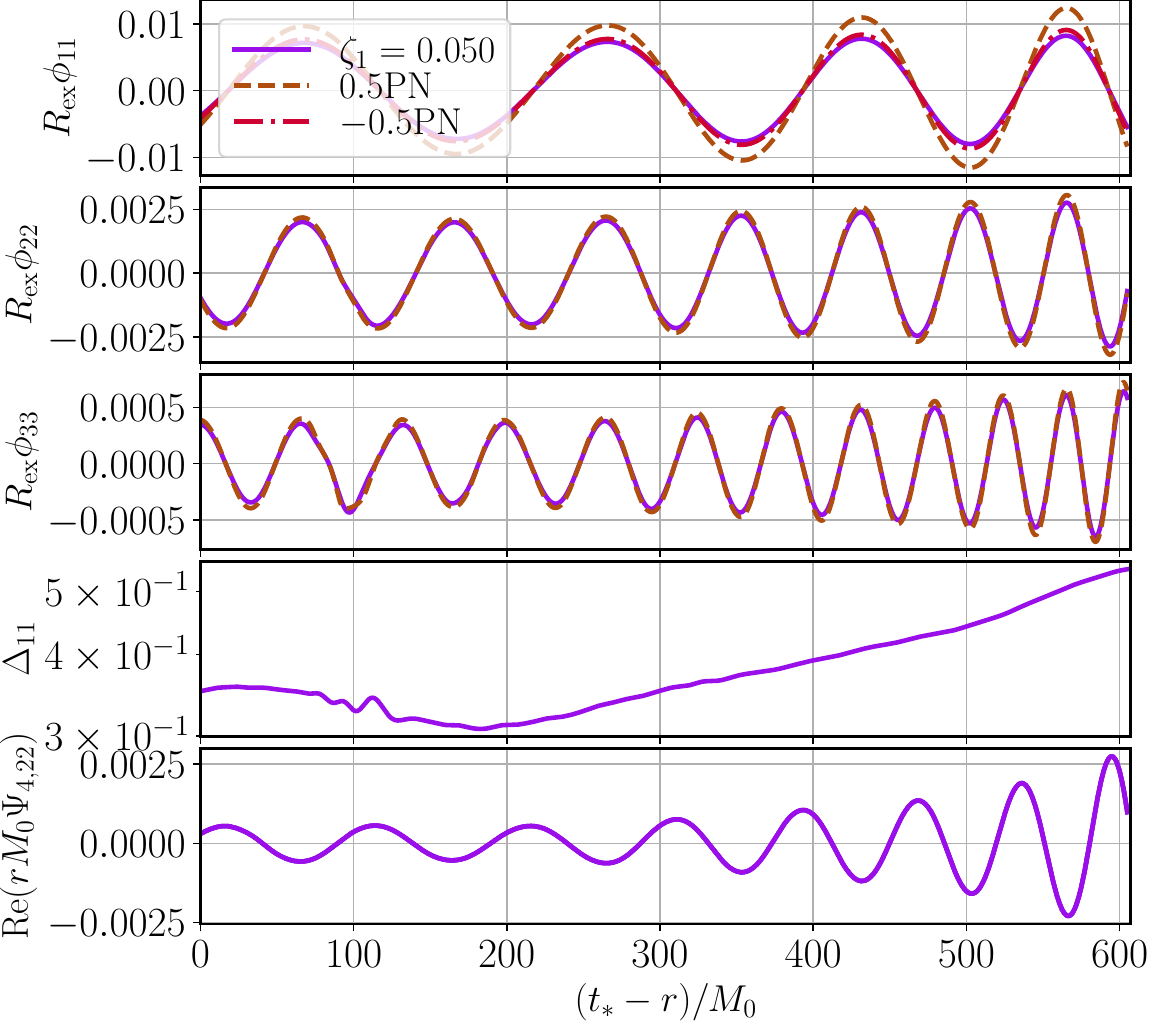}
      \caption{$q=1/2$}
   \end{subfigure}
\caption{
	Scalar waveforms as a function of retarded time, $t_{*} - r = t - t_{\rm align} -r $, 
	rescaled by the extraction radius 
	$R_{\rm ex}=r/M_0 = 90$,
	sourced by nonspinning BH binaries of mass ratio $q=\{1,2/3,1/2\}$ (clockwise from the top left).
        The corresponding waveform $\Psi_{4,22}$ is displayed in the bottom for comparison. 
	We show the $(\ell,m)=(2,2)$ and $(4,4)$ spherical harmonic components for the equal mass ratio and 
	the $(\ell,m)=(1,1)$, $(2,2)$, and $(3,3)$ components for unequal mass ratios. During the inspiral, we also
	display the PN waveform (brown dashed lines), derived to 0.5PN order,
	and the leading order waveform at -0.5PN for the $(\ell,m)=(1,1)$ mode
	(red dash-dotted lines).
	We also show the relative difference between the amplitude of the PN and numerical waveform 
    $\Delta_{lm}$ for the leading order mode.
\label{fig:comparison_sf_pn}
}
\end{figure*}

In Fig.~\ref{fig:comparison_sf_pn}, we compare the leading order scalar
waveforms $\phi_{\ell m}$ from our numerical evolution to the PN formulas
given in Eq.~\eqref{eq:pn_formulas_integrated_scalar_field}.
The PN formulas are accurate to $0.5 \rm PN$ order for the mode
$\phi_{11}$, and to leading $\rm PN$ order for the remaining modes.
As in the comparisons of scalar waveforms computed in 
Refs.~\cite{Witek:2018dmd,Shiralilou:2020gah,Shiralilou:2021mfl},
the frequency we use in the PN expressions 
are obtained from our numerical evolutions
using Eq.~\eqref{eq:orbital_frequency}, 
so our comparison is measuring the accuracy of the PN
approximation in determining the amplitude of the scalar field, given
its frequency.
We see that the fractional difference between the $0.5$PN order PN theory for the $\ell=1,m=1$
mode, and the numerical scalar waveform is about 30$\%$ initially, 
and grows as the binary inspirals.
We also note that the inclusion
of higher PN terms increases the overall amplitude of the scalar waveforms, making
the agreement between the PN and numerical waveforms worse than at leading order at
the frequencies we consider. 
This result holds for all three mass ratios we considered.
Comparing other values of the coupling constant shows
similar behaviour and thus, we do not show the plots here.

\begin{figure*}
\centering
   \begin{subfigure}{0.45\textwidth}
      \includegraphics[width=\columnwidth,draft=false]{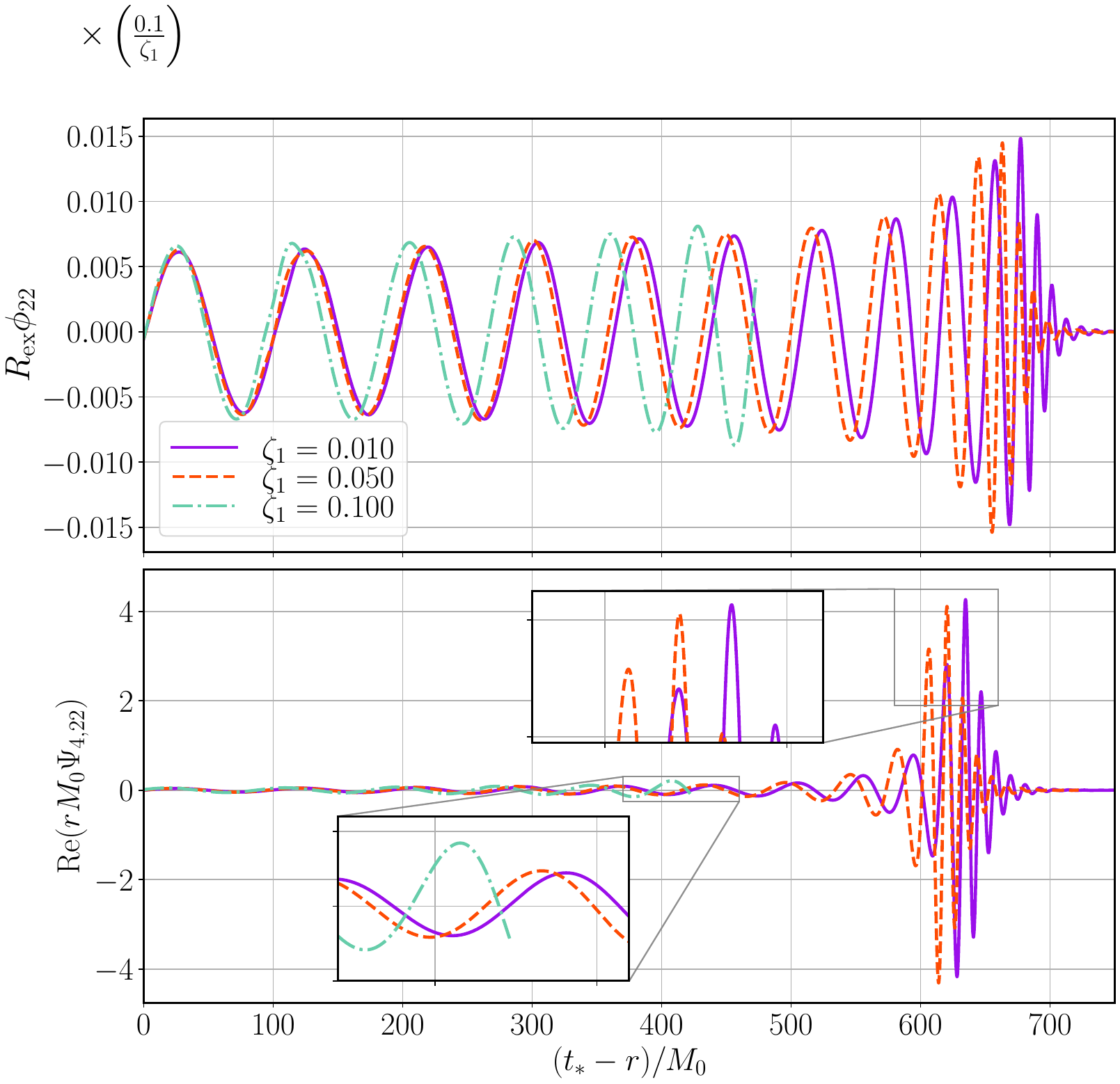}
      \caption{$q=1$}
   \end{subfigure}
   \begin{subfigure}{0.45\textwidth}
      \includegraphics[width=\columnwidth,draft=false]{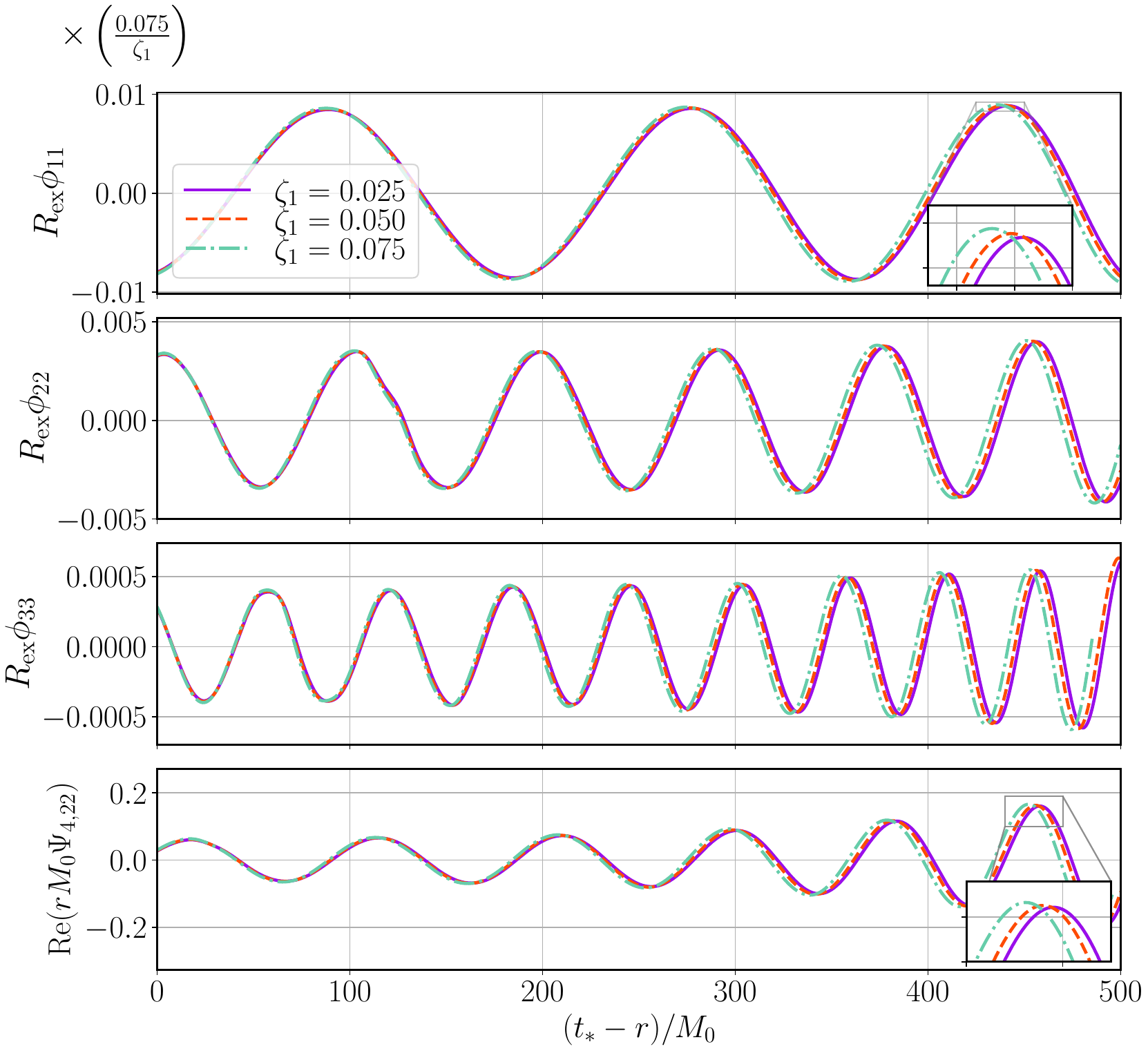}
      \caption{$q=2/3$}
   \end{subfigure}
   \begin{subfigure}{0.45\textwidth}
      \includegraphics[width=\columnwidth,draft=false]{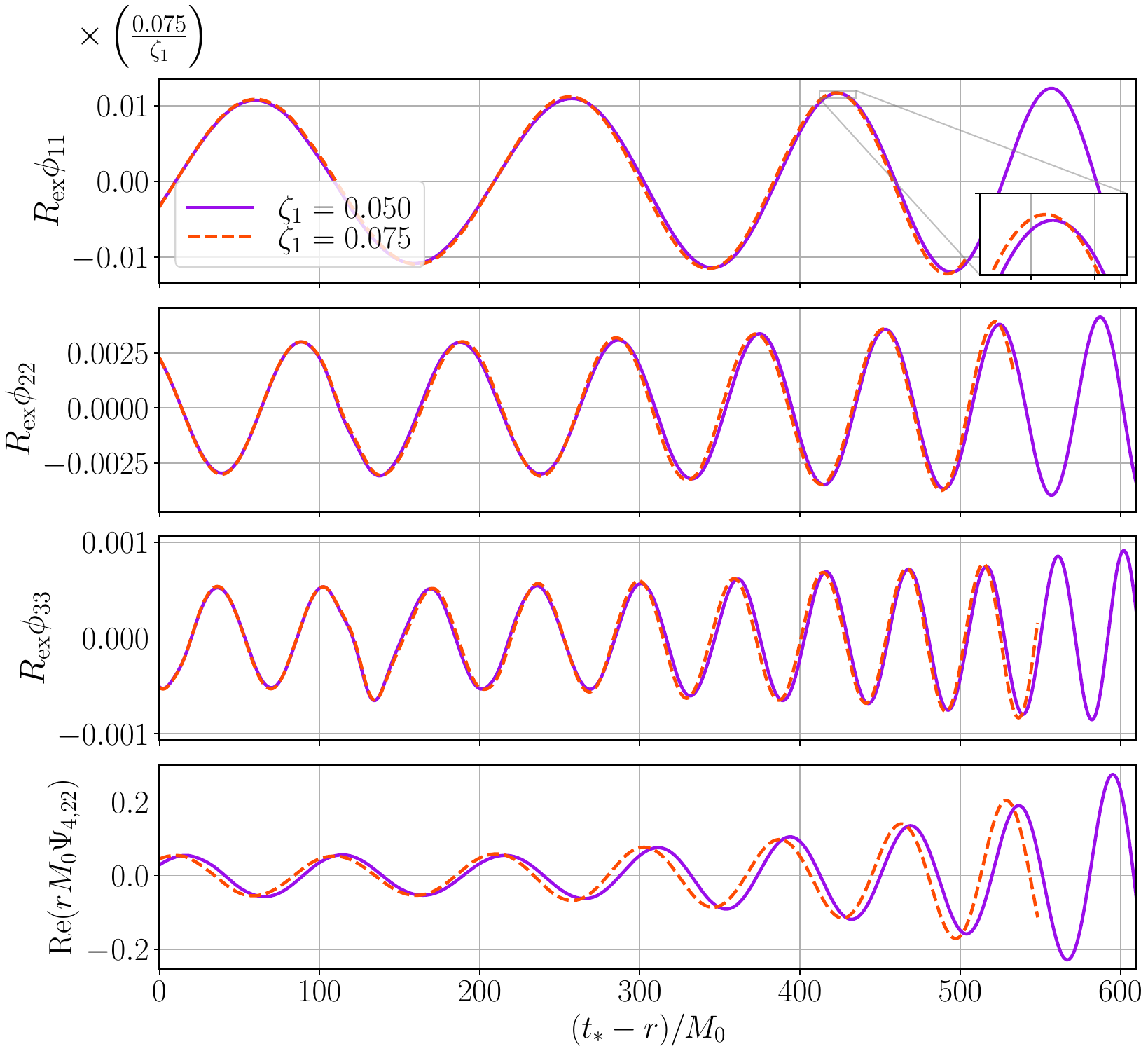}
      \caption{$q=1/2$}
   \end{subfigure}
\caption{
	Scalar waveforms as a function of retarded time,$t_{*} - r = t - t_{\rm align} - r$, 
	rescaled by the extraction radius 
	$R_{\rm ex}=r/M_0 = 90$ and test field dependence on coupling constant $\lambda$,
	sourced by nonspinning binary black holes of mass ratio $q=\{1,2/3,1/2\}$ (clockwise from top left)
	and different coupling constants $\zeta_1$.
        The corresponding gravitational waveforms $\Psi_{4,22}$ are displayed in the bottom 
        of each panel for comparison. 
	We show the leading order $(\ell,m)$ mode for each mass ratio. 
\label{fig:rescaling_nonlinearity_phi}
}
\end{figure*}

Comparing our results to Fig.~7 of Ref.~\cite{Witek:2018dmd}, where the leading order PN scalar
waveforms were compared to numerical waveforms obtained in a test field approximation
(valid to first order in the coupling parameter $\zeta_1$),
we find close agreement between our waveforms, suggesting the test field 
scalar waveform computed from a prescribed orbital evolution
is fairly accurate at least during the early inspiral phase.
This is further emphasized in Fig.~\ref{fig:rescaling_nonlinearity_phi}, 
where we plot the scalar waveforms,
rescaled by $\zeta_1$. In the decoupling limit, the amplitude of the
emitted waveforms is directly proportional to $\zeta_1$ 
\cite{Yagi:2011xp,Witek:2018dmd}.
From Fig.~\ref{fig:rescaling_nonlinearity_phi}, we see that, at least
during the inspiral phase of binary black hole evolution, this relation
holds up well for the full theory. 
This is to be expected, as nonlinear corrections to $\phi$ only
enter at order $\zeta_1^3$ in sGB gravity; 
see Appendix~\ref{sec:perturbative_method}.

\begin{figure*}
\centering
   \begin{subfigure}{0.45\textwidth}
      \includegraphics[width=\columnwidth,draft=false]{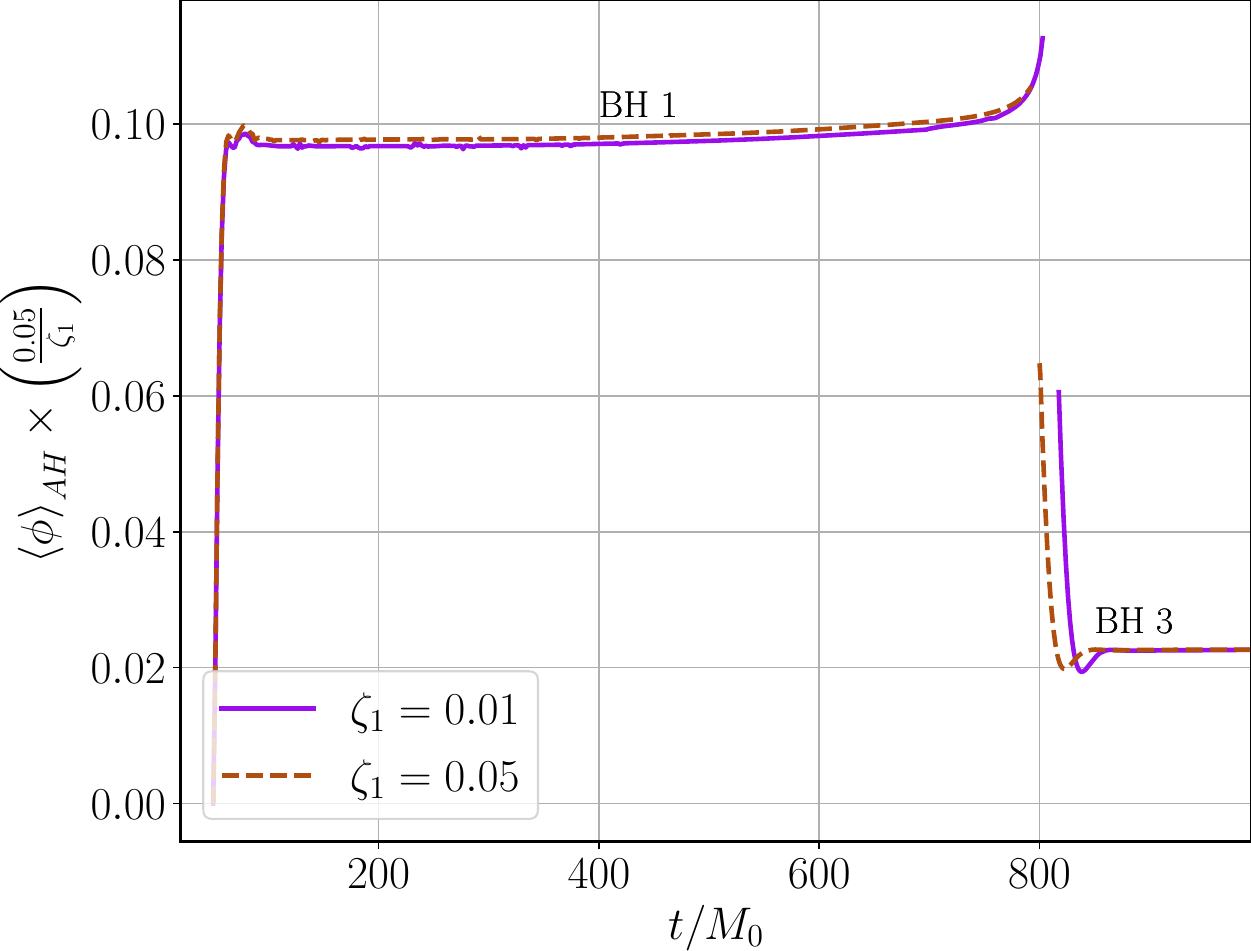}
      \caption{$q=1,\zeta_1=0.01$ and $0.05$}
   \end{subfigure}
   \begin{subfigure}{0.45\textwidth}
      \includegraphics[width=\columnwidth,draft=false]{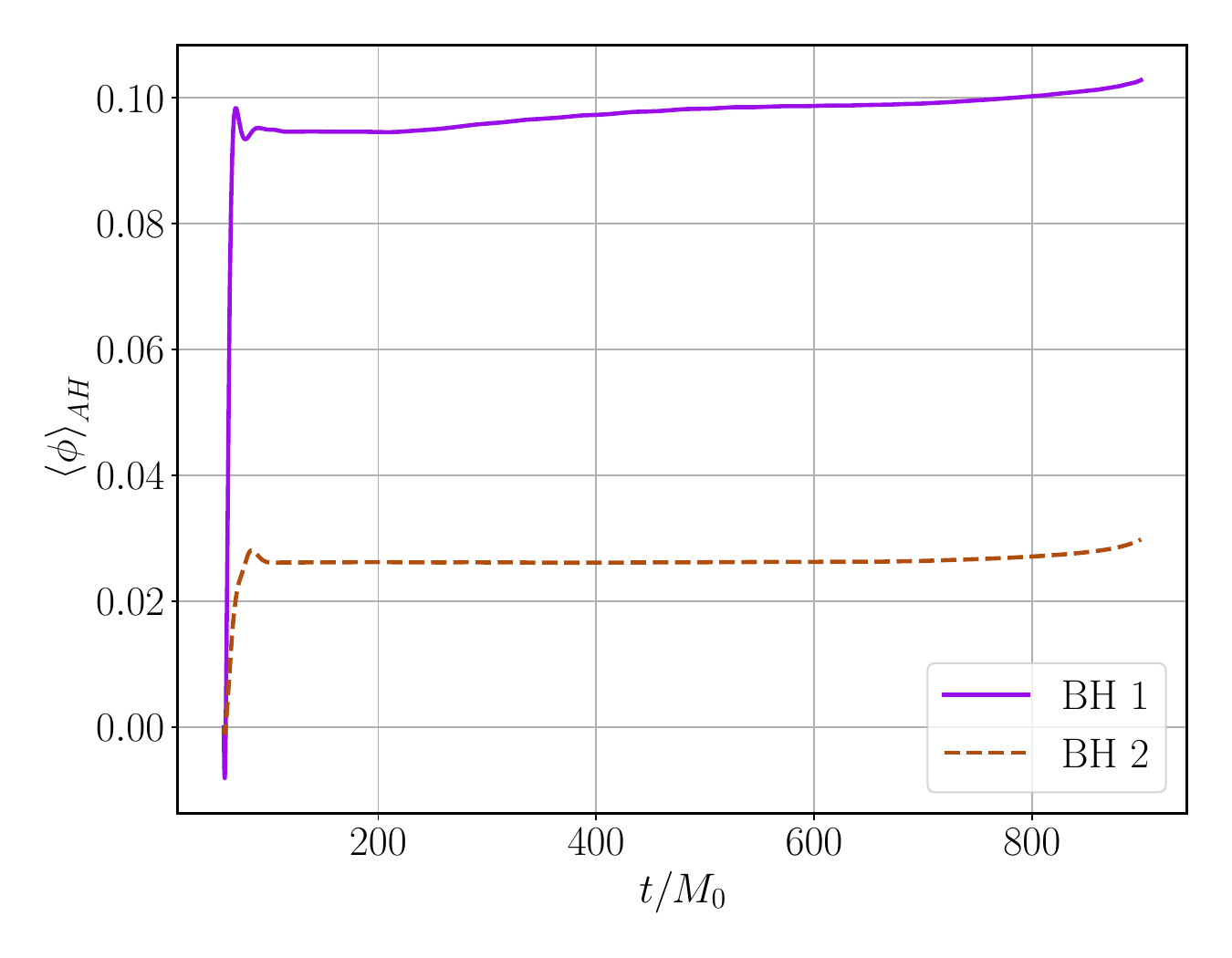}
      \caption{$q=1/2,\zeta_1=0.05$}
   \end{subfigure}
\caption{Average value of the scalar field, rescaled by the test field dependence,
	over the black hole horizons for different mass
   ratios. For the equal mass ratio binary (left panel), we were able to evolve through
   merger, and thus determine the average value of the scalar field on the third, 
   remnant black hole.
   While we were unable to evolve through merger for the unequal mass ratio binaries,
   on the right panel we show the average scalar field for a $q=1/2$ run.
   The dips in the average scalar field near the end of the evolution for that run
   are due to numerical error.
   \label{fig:horizon_phi}
}
\end{figure*}

In Fig.~\ref{fig:horizon_phi}, we plot the average value of $\phi$
on the black hole apparent horizon for the two initial black holes, and
the final remnant black hole, for runs with $(q=1,\zeta_1=0.01$ and $0.05)$ 
and $(q=1/2,\zeta_1=0.05)$.
We see that after the black holes have acquired a scalar charge,
the average value of the scalar field on the two black hole horizons increases
as they inspiral towards each other, in general qualitative agreement
with the predictions of Refs.~\cite{Julie:2019sab,Julie:2022huo}.
The remnant black hole for the equal mass runs (BH3 in the left panel)
has a smaller average scalar field value on its horizon than the two original black holes, 
as it has a larger mass
(so $\lambda/m_3^2 < \lambda/m_{1,2}^2$), 
and it is spinning \cite{East:2020hgw}.
As we discuss in Sec.~\ref{sec:merger}, 
we are unable to evolve through merger for any of 
the unequal mass ratio cases we consider, so there is no remnant apparent horizon in the right panel.

\subsection{Gravitational Waves}
\begin{figure*}
\includegraphics[width=\columnwidth,draft=false]{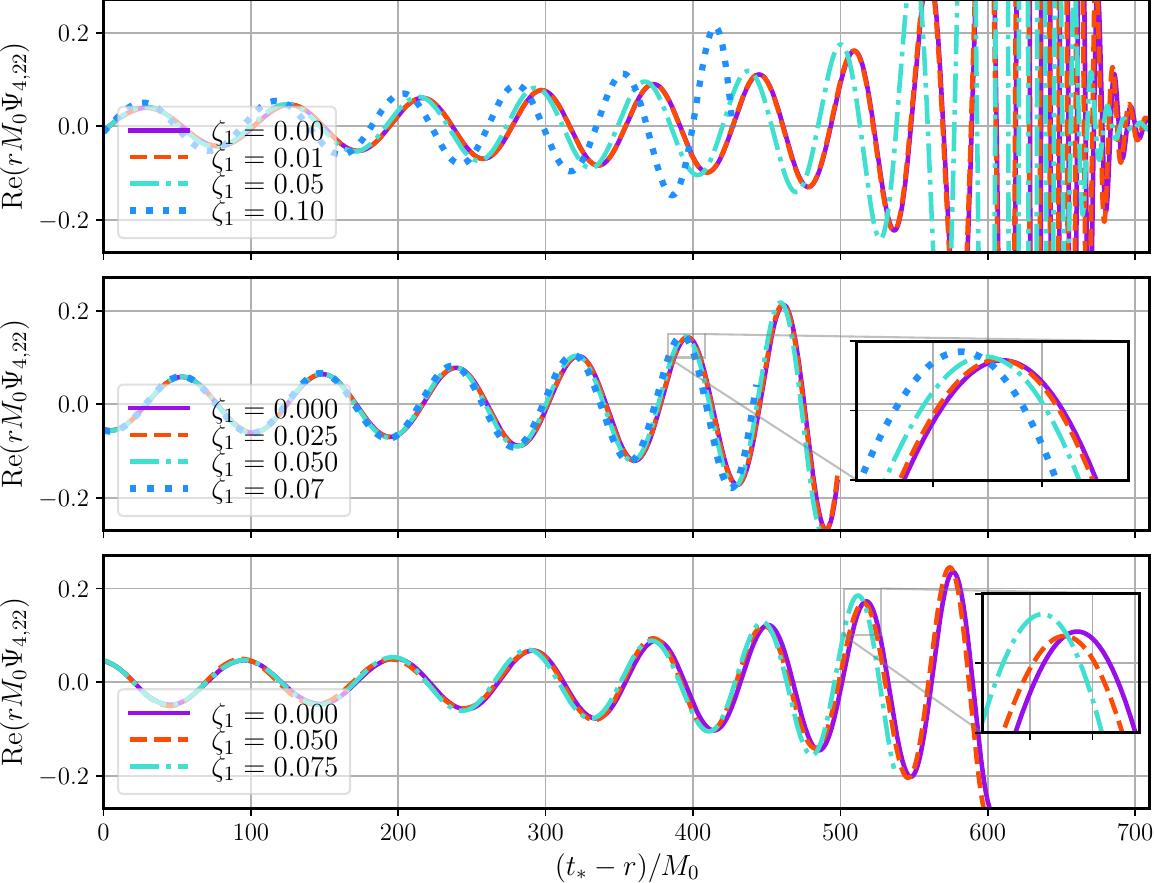}
\caption{
     The radially rescaled value of $\Psi_{4,22}$ 
	as a function of retarded time, $t_{*} - r = t - t_{\rm align} - r$, for different values of $\zeta_1$. 
     The top, middle, and bottom panels show the waveforms for the
     $q=1$, $2/3$, and $1/2$ mass ratio binaries. 
     Here we measure $\Psi_{4,22}$ at a radius of $R_{\rm ex}=r/M_0 = 90$.
     \label{fig:dephasing_psi4}}
\end{figure*}

\begin{figure*}
\includegraphics[width=\columnwidth,draft=false]{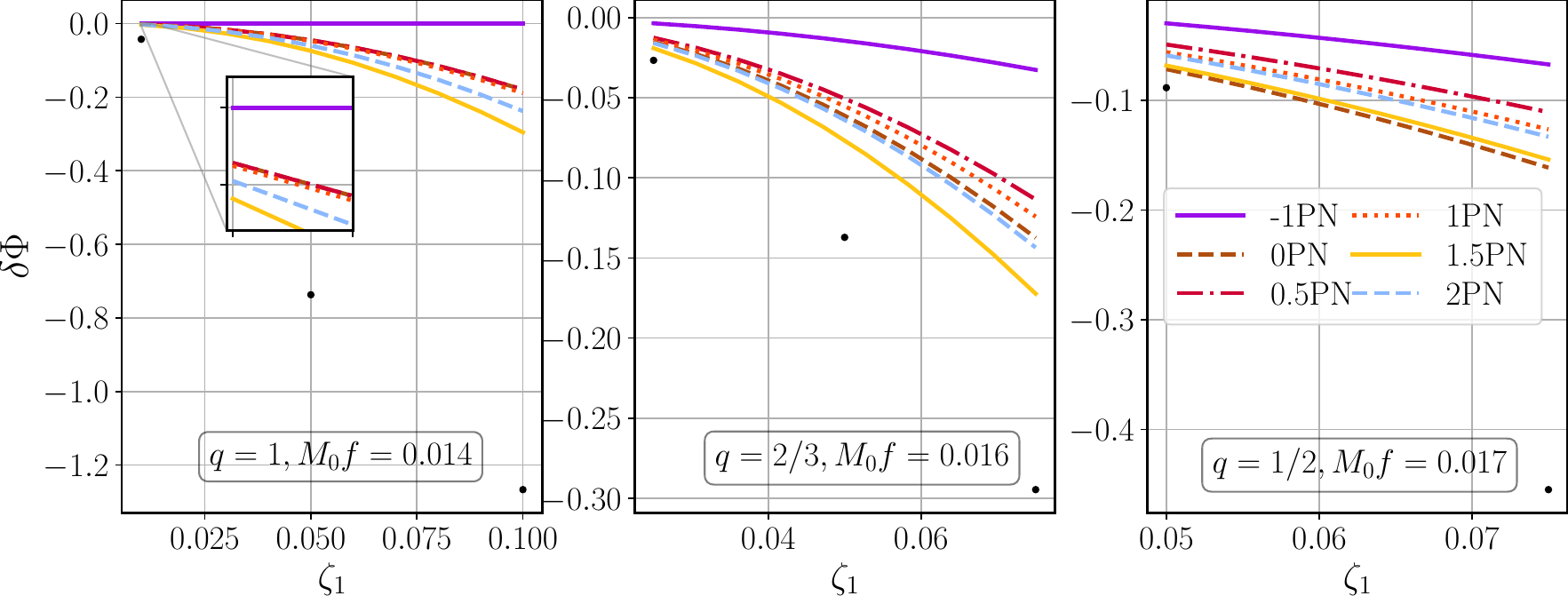}
	\caption{Difference between the orbital phase of gravitational waveform
	in sGB and GR, $\delta \Phi$ [see Eq.~\eqref{eq:def_dephasing}],
    accumulated as the binary evolves from a frequency $f_0=0.01/M_0$ to
        a frequency $f$.
	The left, middle, and right panels display results for the $q=1$, $2/3$, and $1/2$
        mass ratio binaries, respectively, with $M_0f=0.014$, 0.016, and 0.017. We plot the PN predictions for
        orders $-1$PN through $2$PN (with each curve including all terms up to that order).
        }
\label{fig:dphi_freq}
\end{figure*}

\begin{figure*}
\includegraphics[width=\columnwidth,draft=false]{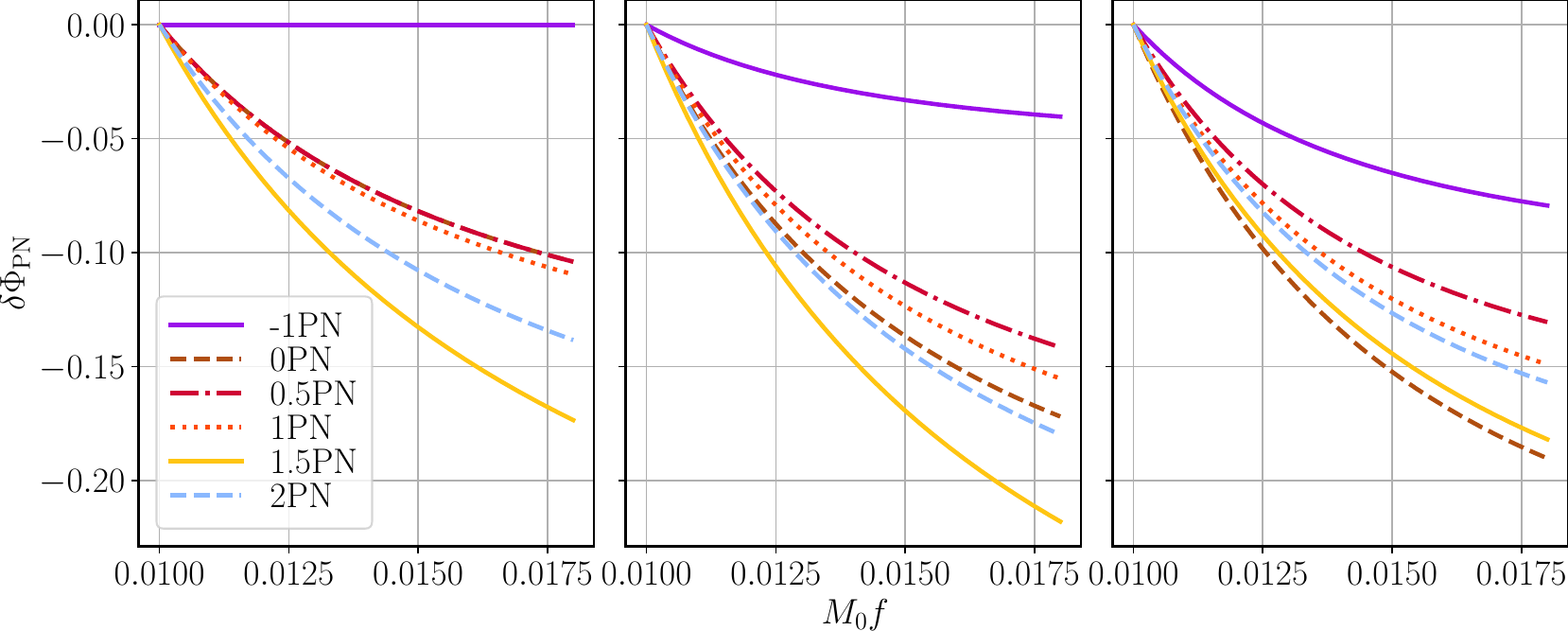}
\caption{The shift in the gravitational wave phase for the orbital phase, summed
        to each PN order up to $2$PN.
        In the left, middle, and right panels we set
        $\zeta_1=0.05$, 0.075, and 0.075, respectively.
	As in Fig.~\ref{fig:dephasing_psi4}, the left panel is for a 
        $q=1$ binary, the middle panel is for a $q=2/3$ binary, and the right
        panel is for a $q=1/2$ binary.}
\label{fig:pn_vs_freq}
\end{figure*}

We next estimate the relative dephasing of the
gravitational waveforms, taking into account various sources of numerical error in
our simulations.
Accurately computing the phase of a gravitational signal is crucial, 
given this will be the most salient effect of sGB gravity
that current gravitational wave detectors are able to measure 
\cite{Yunes:2016jcc,Berti:2018cxi,Lyu:2022gdr}.
Due to the presence of scalar charge around each black hole in sGB gravity, 
black holes will emit scalar radiation as they inspiral each other, 
so they will inspiral faster as compared to what would be the case in GR. 
In Fig.~\ref{fig:dephasing_psi4}, we plot the gravitational waveforms  
$\Psi_{4,22}$, after matching their frequency at a time $t_{\rm align}$, and
applying a rotation in the complex plane, so that their phases align initially. 
We see that there is a noticeable dephasing of binaries with different values of $\zeta_1$.
In Fig.~\ref{fig:dphi_freq}, we quantify the dephasing for the $\ell=2,\ m=2$
mode of $\Psi_4$ [see Eq.~\eqref{eq:psi4_22}]
\begin{align}
   \label{eq:dephasing_formula}
   \delta \Phi(f) 
   \equiv 
   \Phi_{{\rm sGB}}(f) 
   - 
   \Phi_{{\rm GR}}(f)
   ,
\end{align}
by comparing the orbital phase
[computed from Eq.~\eqref{eq:psi4_22}]
of the waveforms
at a given frequency smaller than $M_0 f < 0.018$ which corresponds to the empirically 
found transition from the inspiral to merger-ringdown phase in GR
\cite{Husa:2015iqa,Khan:2015jqa}, along with the corresponding PN predictions
for a quadrupolar driven inspiral
\cite{Sennett:2016klh,Lyu:2022gdr} 
(see also Appendix~\ref{sec:pn_theory}). 
We find $\delta \Phi < 0$, and the dephasing
grows as we increase the coupling $\lambda$, which is in general qualitative
agreement with PN predictions for sGB gravity.
This being said, 
at least for the last few orbits of the inspiral that we study, 
we find that our results do not agree quantitatively with PN predictions. 
A possible reason for this is because we are comparing to PN theory close to the
merger phase of binary evolution, where more orders of the PN expansion are needed
to match to numerical relativity simulations even in GR.
These differences also need to be compared to the various sources of numerical
error in the simulations, which in some cases exceed the small phase differences, as we discuss below.
In Fig.~\ref{fig:pn_vs_freq}, 
we show the dephasing at consecutive orders up to 2PN for
for a range of gravitational wave frequencies we sample in our simulations
(the last few orbits before merger),
yet within the regime where the PN approximation should be valid in GR, $M_0 f < 0.018$
\cite{Husa:2015iqa,Khan:2015jqa}.
The PN formulas we plot were first 
presented including terms of up to 2PN order in Ref.~\cite{Lyu:2022gdr};
we review their computation in Appendix~\ref{sec:pn_theory}.
As noted in Ref.~\cite{Lyu:2022gdr}, we mention that
the dephasing for ESGB gravity has only been
computed to 2PN order, with only partial results at $0.5$ PN order onwards.
We see that there are still noticeable differences in the PN approximation 
with the addition of the highest order terms
in the near-merger regime studied here, and thus the expansion will likely have to be continued to higher
order to achieve a highly accurate prediction in that regime,
although we cannot rule out that the inclusion of the currently missing terms to the $0.5$ through 
$2$ PN
contributions in the phase may lead to a faster convergence in the PN expansion
than observed here.

Finally, we compare the orbital dephasing to the numerical errors in the simulations.
A detailed error analysis is given in Appendix~\ref{sec:error_analysis}, which we briefly summarize
here.
The error in the Richardson extrapolated phase is $\sim 0.25$ radians, which 
is comparable to the ESGB dephasing, and
larger than the relative
error in the $2$PN computation.
However, if the dominant truncation error in our simulations
does not depend strongly on the value of $\zeta_1$, and thus 
partially cancels out when calculating the difference $\delta \Phi$ in the phase
between the sGB and GR simulations using the same resolution, this will lead to noticeable smaller
truncation error in this quantity compared to the overall phase.
We see evidence that this is the case, 
for example, by comparing a measure of the truncation error in $\delta \Phi$, computed by comparing
a $q=1/2$ GR simulation to an equivalent sGB simulation with $\zeta_1=0.075$ at two different resolutions,
to an estimate of the overall truncation error in $\Phi$ for the same sGB case. 
We find the former to be $\sim50\times$ smaller than the latter (see Appendix~\ref{sec:error_analysis}).
We also find similar results for the GW amplitude.
Thus, for a number of cases (see Fig.~\ref{fig:dphi_freq}), 
the difference in errors
is smaller than the dephasing $\delta\Phi$ we measure.

Lastly, we note that
the dephasing between the sGB and GR simulations may be caused
by small differences in the eccentricity of our simulations, which would
be caused by the orbit being slightly perturbed by the rapid development of the scalar field
around the black holes at early times, as an artifact of using initial conditions with $\phi=\partial_t \phi=0$.
If this were the case, one would expect the eccentricity of 
the modified waveforms to increase with coupling. 
We estimate the orbital eccentricity in our simulations to be $\lesssim 0.01$, 
and we find that it decreases with increasing resolution, with only a mild dependence on
coupling. This suggests that residual eccentricity from the initial data
is subdominant to finite-resolution numerical errors, and does not significantly
affect the dephasing of the binary.
The eccentricity of the binary system is not much affected by the value
of the Gauss-Bonnet coupling,
as even for the largest couplings we consider 
the energy contained in the scalar cloud is only a small fraction of the total 
binary binding energy,
and an even smaller fraction of that energy 
is radiated away during the scalarization process.

\subsection{Merger dynamics}
\begin{figure*}
\includegraphics[width=0.49\columnwidth,draft=false]{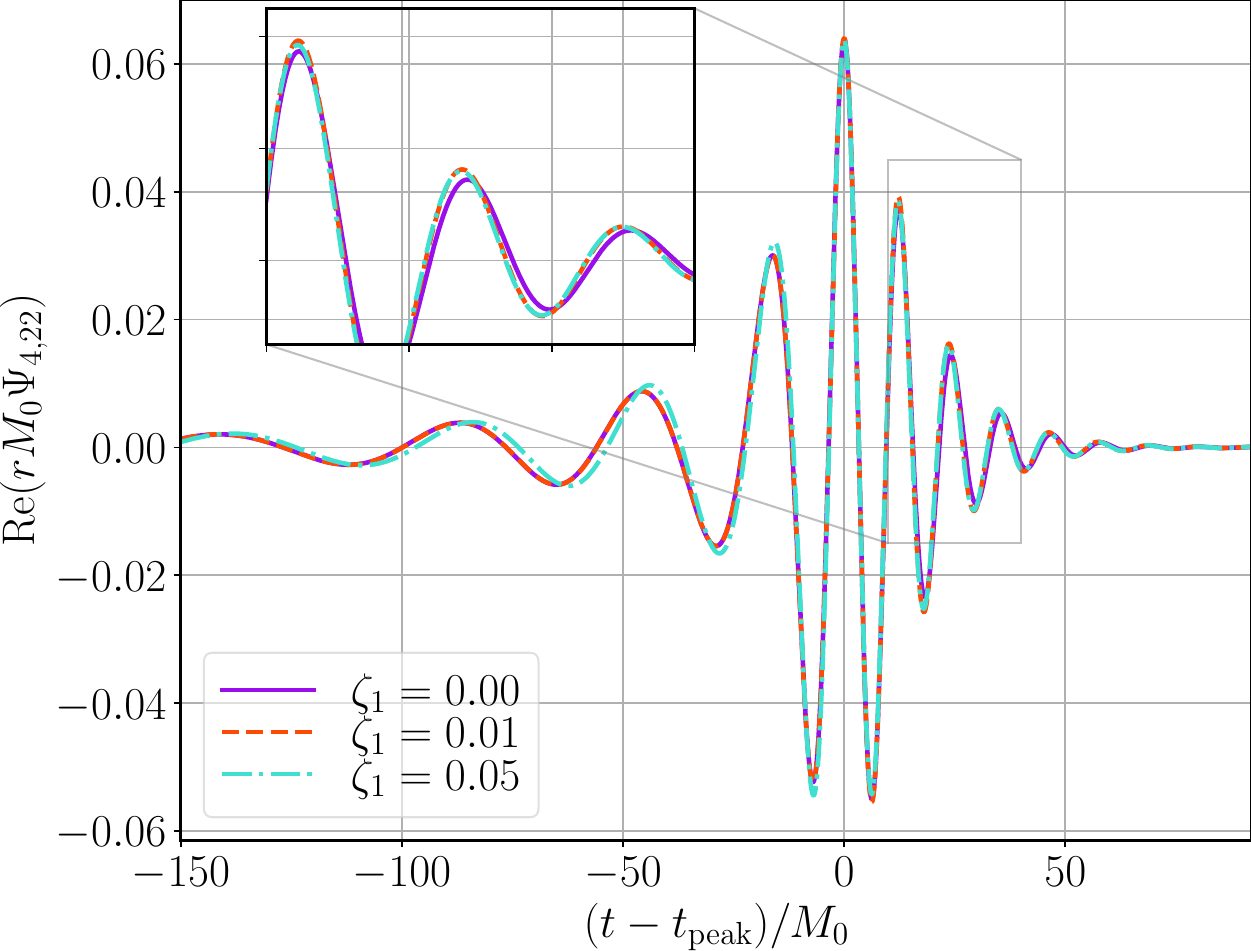}
\includegraphics[width=0.49\columnwidth,draft=false]{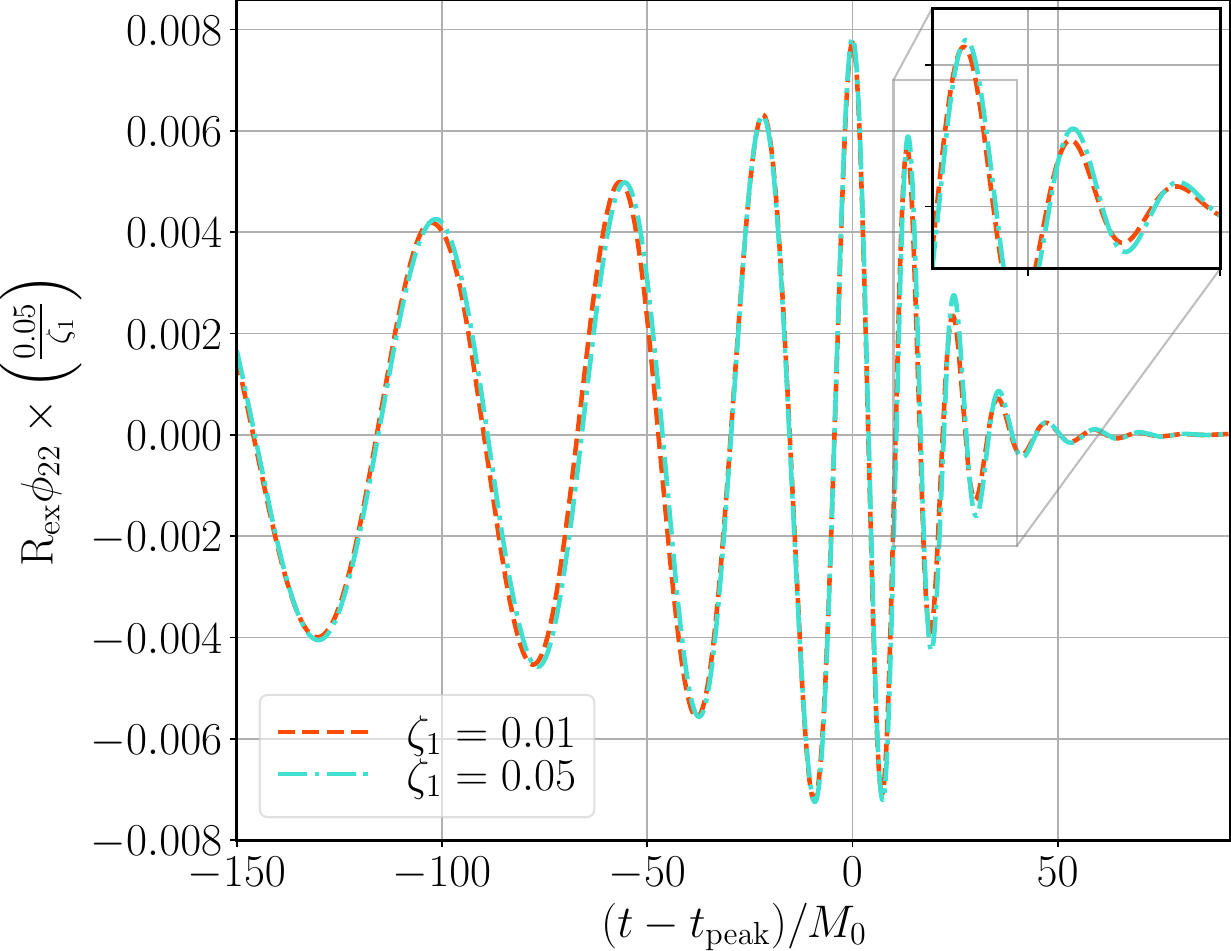}
	\caption{Gravitational wave radiation (left) and scalar radiation (right) for
	equal mass ratio binaries with coupling $\zeta_1 = 0$, 0.01, and $0.05$. 
	We show the real part of the $\ell = m =2$ spherical harmonics of the
	Newman-Penrose scalar $\Psi_{4}$ and $\phi$. Time is measured
	with respect to the time where the complex amplitude of $\Psi_{4,22}$/$\phi_{22}$
	peaks. We add an overall phase so that the waveforms are real and positive at 
	$t=t_{\rm peak}$.}
\label{fig:merger}
\end{figure*}

Lastly, we mention the effects of ESGB on the merger dynamics of equal mass binaries
with couplings $\zeta_1 = 0.01 $ and $0.05$, compared to GR. 
Figure~\ref{fig:merger}
shows the gravitational wave emission starting slightly before merger,
and including the ringdown, for different values of $\zeta_1$. 
We find that while the ESGB
waveforms have a noticeable dephasing relative to GR, consistent
with the fact that ESGB binaries should merge faster 
due to the additional energy loss through scalar radiation,
the peak amplitude of the gravitational wave at merger depends only
very weakly on $\zeta_1$.
The effect of modified gravity on the frequency and decay rate of the quasinormal modes is 
also too small to reliably quantify with our current numerical data, 
so we defer a more detailed study of the ringdown to future work.

In the right panel of Fig.~\ref{fig:merger} we show the leading $\ell =m=2$
mode of the scalar waveform after rescaling for the test-field dependence on
the coupling, which implies that the amplitude of $\phi$ scales linearly with
$\zeta_1$.  For the $\zeta_1 =0.05$ case, we find an additional nonlinear
enhancement in the scalar field amplitude at merger, with $|\phi|/\lambda$
roughly  {$5 \%$} higher compared to the $\zeta_1 =0.01$ case. 

The negligible effect on the GW amplitude with varying ESGB coupling
that we find here contrasts with the large effect found in order-reduced
simulations. In particular, the correction to $\Psi_4$,
which scales quadratically with $\zeta_1$ in the perturbative approach taken 
in Ref.~\cite{Okounkova:2020rqw}, gives an order-one correction to the amplitude 
for the highest couplings used here (see Fig.~2 of Ref.~\cite{Okounkova:2020rqw});
though we note that Ref.~\cite{Okounkova:2020rqw} also uses a slightly different
mass-ratio ($q=0.82$) and non-zero spins for the constituent black holes.
We speculate that this qualitative difference behavior in the waveform is
due to the presence of secularly growing errors terms, which are known
to be present in such a perturbative approach to evolving modifications
to GR. For more discussion of this phenomena, 
see Refs.~\cite{Okounkova:2019dfo,Okounkova:2020rqw,GalvezGhersi:2021sxs}.

%=============================================================================
\section{Discussion and Conclusion}
\label{sec:conclusion}
In this work, we have performed the first systematic study of the nonlinear
dynamics of binary black hole inspiral and merger in sGB gravity. 
We considered several values of the sGB coupling and the binary mass
ratio, and compared our results to PN theory.
Solving the full equations of motion allowed us to directly measure
the increased dephasing of the inspiral due to the emission of 
scalar radiation, and to determine the relative effects of nonlinearity
on the scalar and gravitational waveforms.
We argue that, 
at least in the last few orbits
of the inspiral phase before merger, PN theory is currently not accurate
enough to determine the dephasing of the binary due to the modified gravity,
even taken as a correction to a more accurate to GR waveform. 

In addition to measuring the dephasing of binary black holes,
we find that leading order PN theory (in the GB coupling $\lambda$)
does well in matching the amplitude of scalar radiation
emitted during the inspiral phase, given the frequency of observed
gravitational radiation. 
This is in general qualitative agreement with earlier numerical relativity
work that compared simulations of sGB gravity in the decoupling
limit to PN predictions \cite{Witek:2018dmd}.
The success of leading order PN theory in matching the scalar waveform can
be partially explained by the fact that
corrections to the scalar field amplitude in the GB coupling enter at
order $\zeta_1^3$ for sGB gravity
(see Appendix~\ref{sec:perturbative_method}).

We have studied the dynamics of the merger for a limited number of cases,
where we found that when the black holes merge the effect 
due to the ESGB modifications on the peak amplitude of the 
gravitational wave signal is small, in contrast
to what results using perturbative treatments of the merger would suggest.
We leave a detailed study of the detectability of these effects
and their degeneracy with different intrinsic parameters
to future work.

In this first study, for computational expediency, and given that 
the ESGB equations of motion
are more complicated to solve than the GR ones, 
we have focused on the roughly last 8 orbits before merger.
However, an obvious direction for future work is to consider binaries that start
at wider separations (and hence lower orbital and gravitational wave frequencies),
in order to determine at what point leading order PN theory becomes accurate.
Modeling the merger is arguably the most important contribution
numerical relativity can make to our understanding of binary
black hole evolution. 
As we were unable to evolve through
merger for larger coupling values
understanding this limitation of our code/methods remains an
important task for future work.
Different choices of gauge or auxiliary metrics, as well developing
better diagnostics for monitoring the breakdown of hyperbolicity
may help address this.
As mentioned above, we believe one of the main difficulties lie in
being able to excise elliptic regions near merger,
around the time
the final remnant black hole forms from the merger.
Our algorithm may be improved by implementing a more complicated
excision surface (currently we only excise an ellipsoidal region),
and working with higher resolution, to more stably excise closer
to the surface of the apparent horizons.
We note that recent work \cite{AresteSalo:2022hua}
reports evolutions of non-spinning,
equal mass ratio black hole binaries 
through merger using a modified CCZ4 formulation of the equations of motion
with puncture-like coordinates, 
for $\zeta_1$ values as large as $\zeta_1=0.1/{\sqrt{2}}\sim0.07$ 
(converting to our conventions).
In that work, the authors make use of an effective excision algorithm
by letting the modified gravity coupling go to zero at small values of the spatial
metric conformal factor, as in Refs.~\cite{Figueras:2020dzx,Figueras:2021abd};
such a method may be useful in conjunction with our direct
excision method to stabilize the evolution near the excision boundary.

In this work, we only considered binary black hole systems where
the individual black holes were initially nonspinning.
As black hole spin can significantly impact the dynamics
of binaries in GR, a natural next step to this work would be to
consider black hole spin. Furthermore, 
introducing spin may lead to novel gravitational
wave signatures as, for example, in
black hole spin-induced spontaneous scalarization 
\cite{Dima:2020yac,Herdeiro:2020wei,Berti:2020kgk}.

We have only simulated the dynamics of arguably the simplest of the ESGB
gravity theories that gives scalar hairy black holes.  Other kinds of scalar
Gauss-Bonnet couplings (i.e. more general terms of the form
$\beta(\phi)\mathcal{G}$ in the action) can allow for a rich range of
phenomena, most notably the effect of spontaneous (de)scalarization, which so
far has only been studied either perturbatively
\cite{Silva:2020omi,Elley:2022ept}, or in symmetry-reduced settings
\cite{Doneva:2017bvd,Silva:2017uqg, Minamitsuji:2018xde,Silva:2018qhn,
Dima:2020yac,Herdeiro:2020wei,Berti:2020kgk,East:2021bqk}.  As well, including
a term of the form $f\left(\phi\right)X^2$ in the action is also ``natural''
from an effective-field theoretic point of view, as this is another four
derivative term that is also
parity-invariant~\cite{Weinberg:2008hq,Kovacs:2020pns}, and may have some
effect on the binary evolution.  Simulating nonlinear effects such as
spontaneous black hole scalarization requires understanding the backreaction of
the scalar field on the background geometry, as that affects the saturation of
the instability and end state black hole, and determines which effects occur in
the regime where the theory remains hyperbolic~\cite{East:2021bqk}.  Accurately
simulating theories with high precision that exhibit spontaneous black hole
scalarization will additionally require the development of initial data solvers
that solve the constraint equations in sGB gravity that have an initially
nontrivial scalar field profile \cite{Kovacs:2021lgk,Ripley:2022cdh}.  It would
also be interesting to extend recent work on binary neutron star
mergers~\cite{East:2022rqi} to study black hole--neutron star binaries in ESGB
gravity (earlier work on spontaneous scalarization in ESGB gravity for single
neutron star solutions include Ref.~\cite{Kuan:2021lol}).

%=============================================================================
\section*{Acknowledgements}
We thank Banafsheh Shiralilou for helpful discussions about
the results in Refs.~\cite{Shiralilou:2020gah,Shiralilou:2021mfl}, and for sharing
unpublished work with us,
and thank Vasileios Paschalidis for sharing a Mathematica notebook that
computes puncture initial data to $2$/$2.5$PN order.
M.C. thanks Nan Jiang and Zhenwei Lyu for several clarifications 
about Ref.~\cite{Lyu:2022gdr}.
J.R. thanks Michalis Agathos, Ulrich Sperhake, Kent Yagi,
and Nicolas Yunes for helpful discussions and correspondence.

M.C. and W.E. acknowledge support from an NSERC Discovery grant.
J.L.R. was supported by STFC Research Grant No. ST/V005669/1. 
This research was supported in part by 
Perimeter Institute for Theoretical Physics. 
Research at Perimeter Institute is supported in
part by the Government of Canada through the
Department of Innovation, Science and Economic Development
Canada and by the Province of Ontario through the
Ministry of Economic Development, Job Creation and
Trade. This research was enabled in part by support
provided by SciNet (www.scinethpc.ca) and the
Digital Research Alliance of Canada (alliancecan.ca). Calculations were
performed on the Symmetry cluster at Perimeter Institute, the Niagara cluster at the University of 
Toronto, and the Narval cluster at Ecole de technologie sup\'erieure 
in Montreal.
This work also made use of the Cambridge Service for Data Driven
Discovery (CSD3), 
part of which is operated by the University of Cambridge Research
Computing on behalf of the STFC DiRAC HPC Facility (www.dirac.ac.uk).
The DiRAC component of CSD3 was funded by BEIS capital funding via STFC capital
grants ST/P002307/1 and ST/R002452/1 and STFC operations grant ST/R00689X/1.

\newpage
%=============================================================================
\appendix
%=============================================================================
\section{\label{sec:error_analysis} Convergence tests 
and accuracy of our simulations 
}
Here we quantify the main sources of error in our simulations, which include
the numerical truncation error, finite radius extraction effects, and residual
orbital eccentricity.

%------------------------------------------------------------------------------
\subsection{Truncation error and convergence}
\begin{figure*}
\centering
   \begin{subfigure}{0.45\textwidth}
      \includegraphics[width=\columnwidth,draft=false]{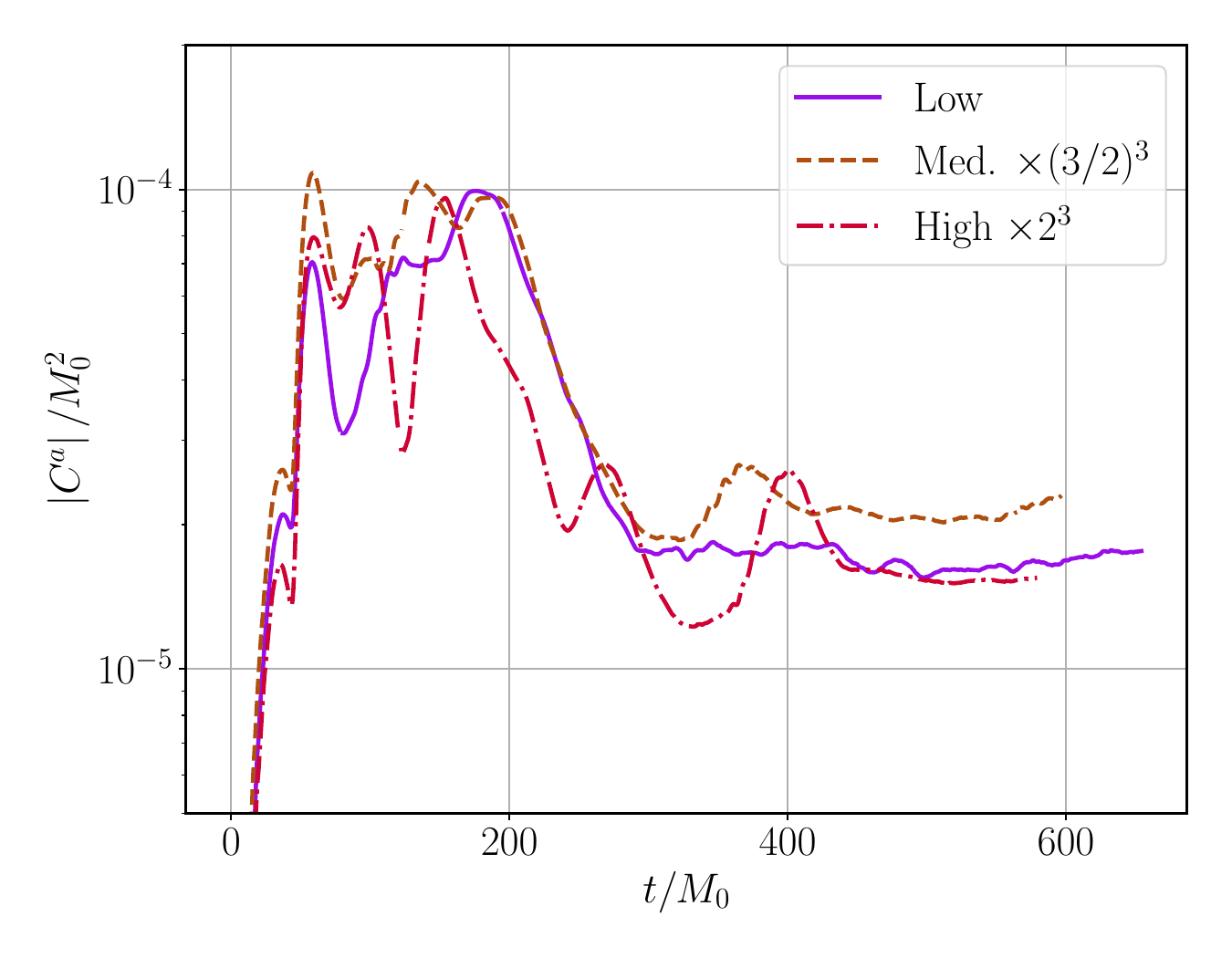}
      \caption{$q=2/3$}
   \end{subfigure}
   \begin{subfigure}{0.45\textwidth}
      \includegraphics[width=\columnwidth,draft=false]{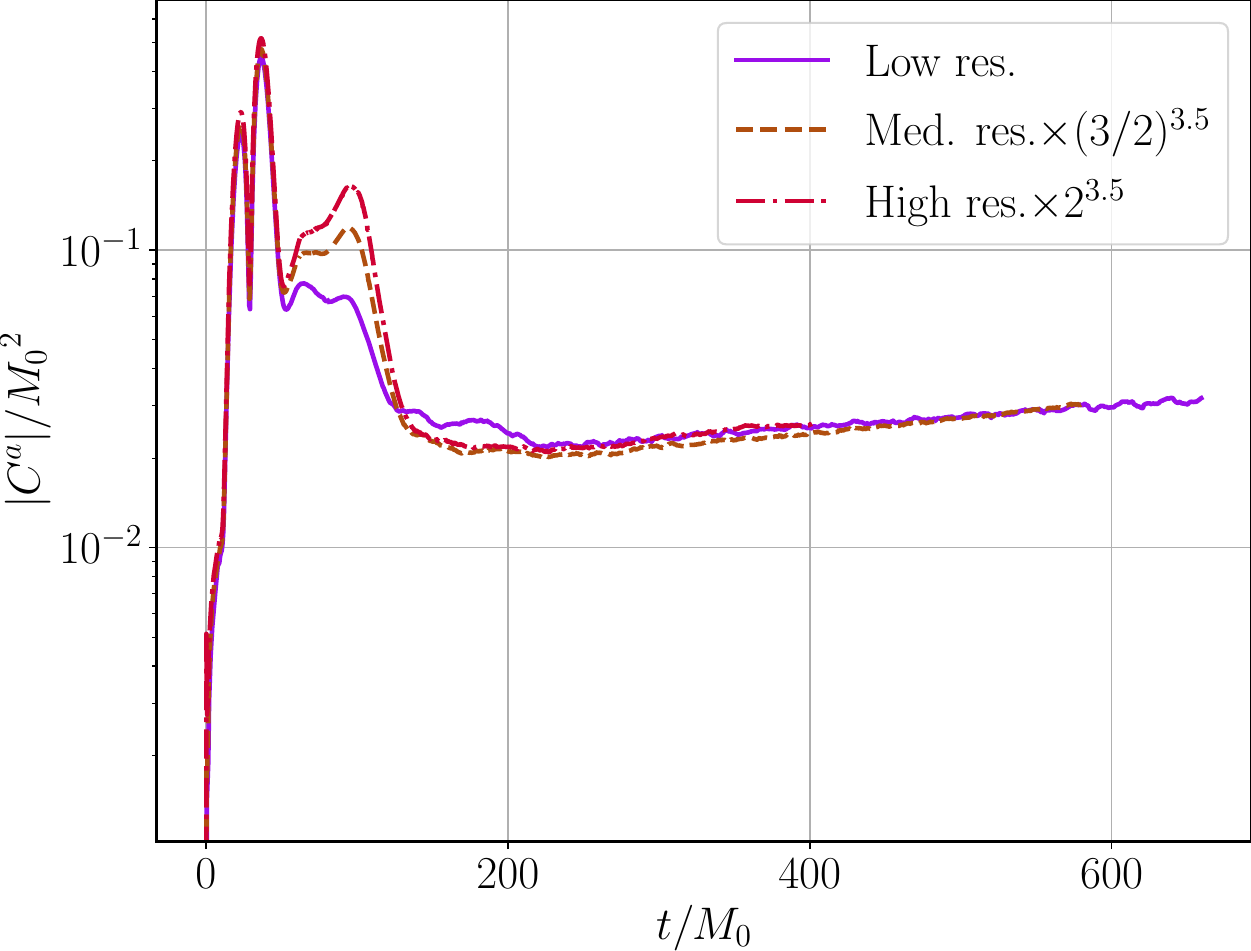}
      \caption{$q=1/2$}
   \end{subfigure}
\caption{
     Volume integrated norm of the constraint violation as a function of time for a nonspinning 
     $q=2/3$ and $q=1/2$
     binary black hole merger with $\zeta_1=0.075$ at three resolutions. 
     The medium and high resolutions have 1.5 $\times$ and $2 \times$ the resolution
     of the low resolution on the coarsest grid.
	We observe roughly third order convergence of our runs, which is
     consistent with the third order in time interpolation used on the
     boundaries of adaptive mesh refinement grids \cite{East:2011aa,PAMR_online}.
\label{fig:cnst_violation}
}
\end{figure*}

We first consider the truncation error, which is due
to the finite resolution of the simulations.
The simulations of the binary black hole systems with mass
ratio $q=1$ and $q=2/3$ presented in this work
use eight levels of adaptive mesh refinement with a refinement ratio
of $2:1$, and have a linear grid spacing of $dx = 0.012 M_0$ on the finest level 
containing the smallest back hole.
The results for the mass ratio $q=1/2$ use nine levels of adaptive
mesh refinement and a grid spacing of $dx = 0.006 M_0$ around the smallest black hole.
In Fig.~\ref{fig:cnst_violation}, we plot the integrated constraint
violation for a $q=2/3$, $\zeta_1=0.075$ binary
with grid spacing that is $4/3$ and $\times 2/3$ as large as default resolution.
We also perform a resolution study of a $q=1/2$, $\zeta_1=0.075$ binary,
where the linear spacing of the medium resolution
is $dx = 0.005 M_0$ and covers the smallest black hole. The integrated constraints 
shown in Fig.~\ref{fig:cnst_violation} have grid spacing $4/3$ and $\times 2/3$ as large
as medium resolution.
We see roughly third order convergence in the constraint violation.
Though we use fourth order finite difference stencils and Runge-Kutta time integration, this level of convergence
is consistent with the third order interpolation in time used to set values on the boundaries
of adaptive mesh refinement levels.

\begin{figure*}
\centering
   \begin{subfigure}{0.45\textwidth}
      \includegraphics[width=\columnwidth,draft=false]{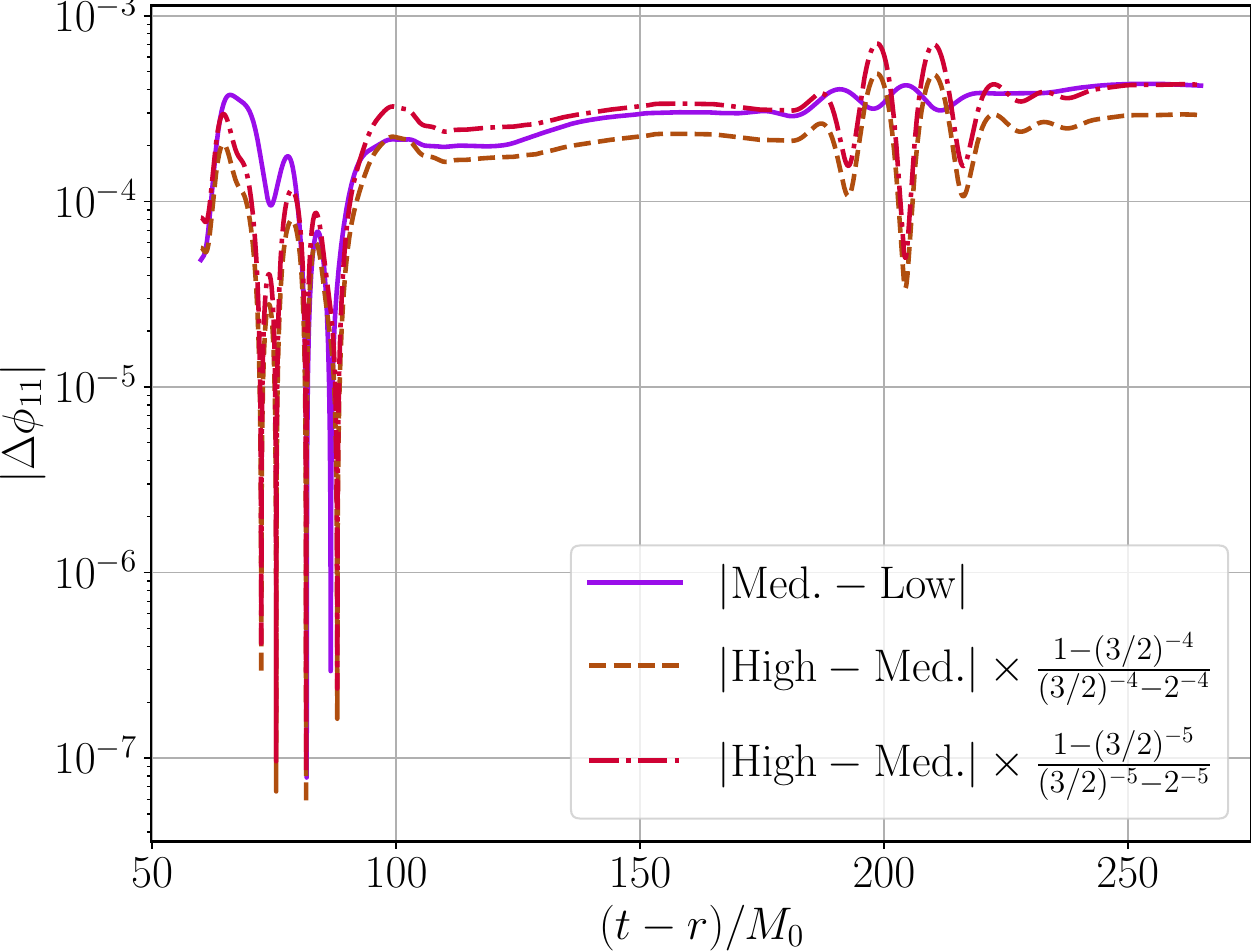}
	   \caption{Amplitude $A$ of $\phi_{11}$ , $q=1/2$}
   \end{subfigure}
   \begin{subfigure}{0.45\textwidth}
      \includegraphics[width=\columnwidth,draft=false]{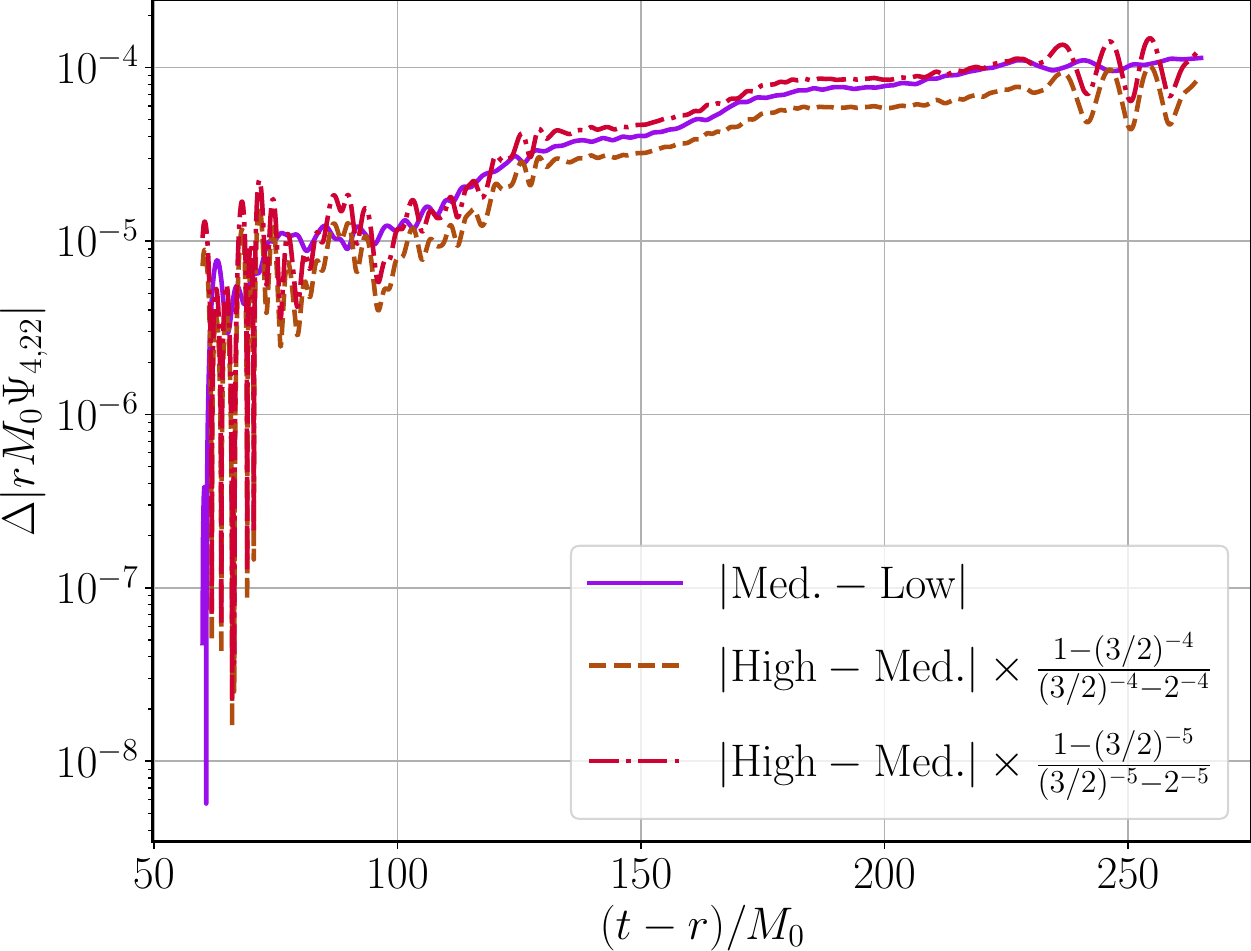}
	   \caption{Amplitude $A$ of $\Psi_{4,22}$, $q=1/2$}
   \end{subfigure}
   \begin{subfigure}{0.45\textwidth}
      \includegraphics[width=\columnwidth,draft=false]{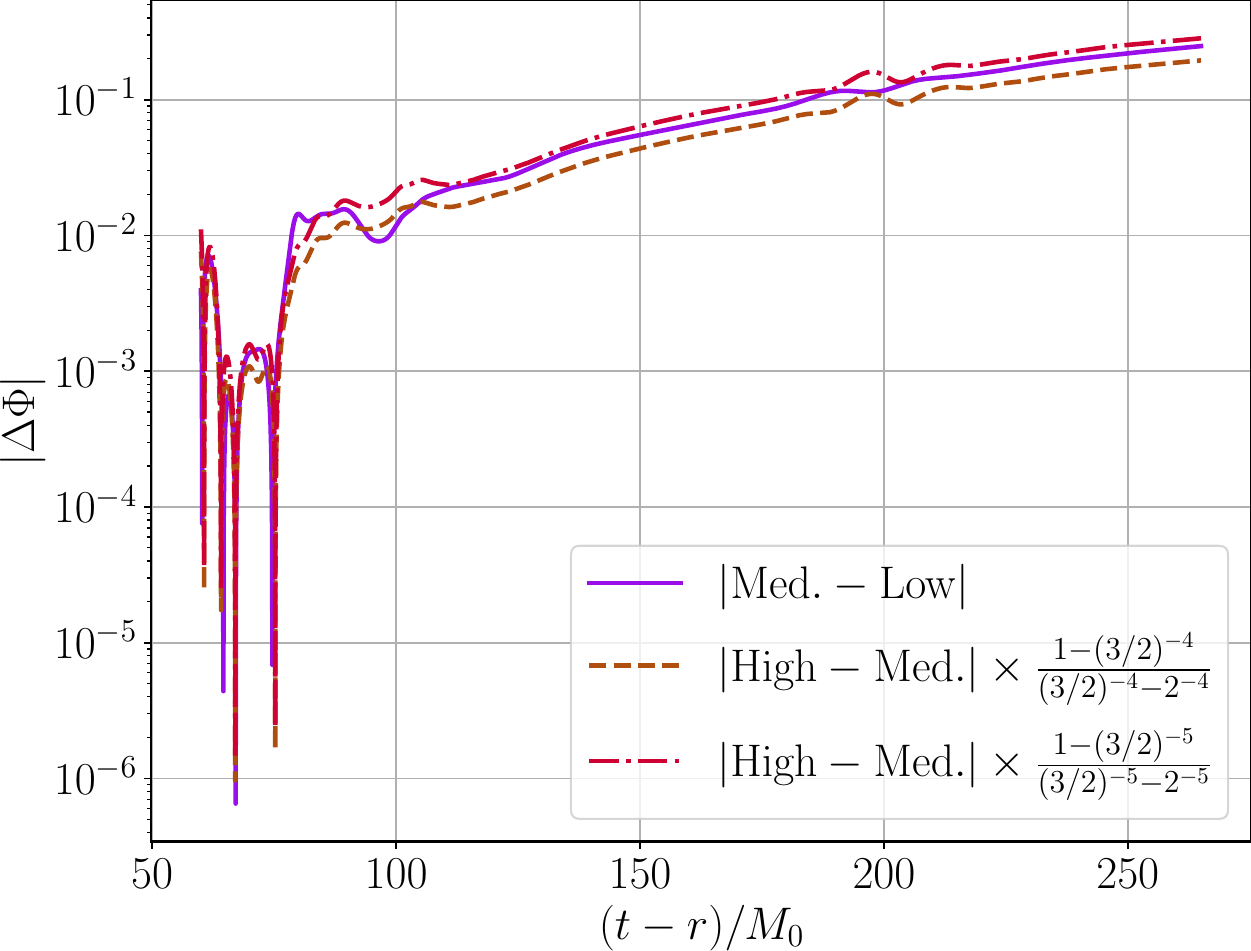}
	   \caption{Phase $\Phi$ of $\phi_{11}$, $q=1/2$}
   \end{subfigure}
   \begin{subfigure}{0.45\textwidth}
      \includegraphics[width=\columnwidth,draft=false]{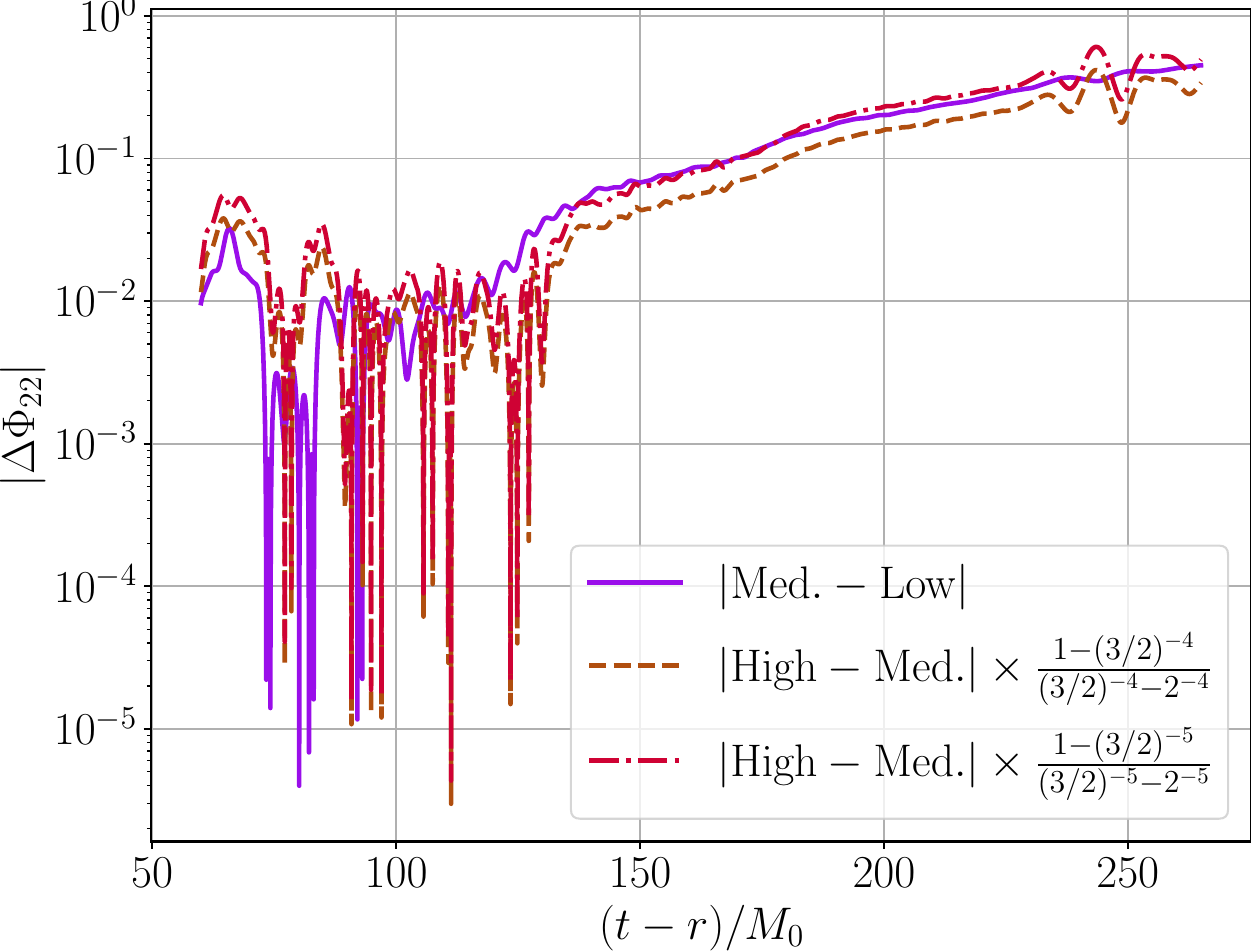}
	   \caption{Phase $\Phi_{22}$ of $\Psi_{4,22}$, $q=1/2$}
   \end{subfigure}
\caption{
	We show the absolute differences between the low, medium, and high resolutions
	of the amplitude and phase of the scalar (left) and tensor (right) waveforms
	for a nonspinning BH binary with mass ratios
	$q=1/2$ and coupling $\zeta_1=0.075$.
    We see that the waveform converges at between fourth order and fifth order (corresponding
    to the scaling used for the dashed and the dashed-dotted lines, respectively).
	Note that we only show the scalar waveform from $50M$ onwards as the scalar field is zero
	before then.
\label{fig:psi4_self_conv_study}
}
\end{figure*}

\begin{figure*}
\centering
   \begin{subfigure}{0.45\textwidth}
      \includegraphics[width=\columnwidth,draft=false]{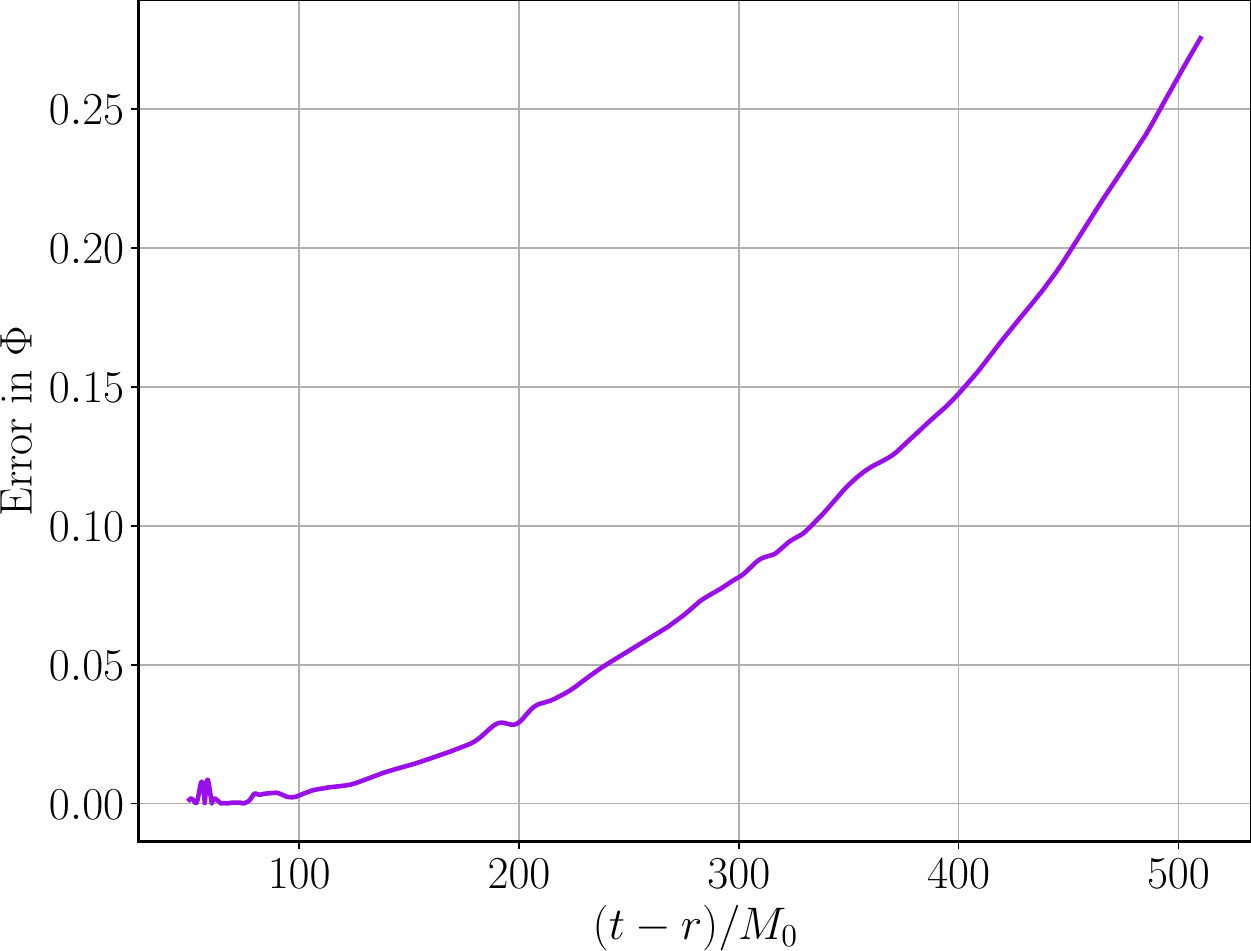}
	   \caption{Phase $\Phi$ of $\phi_{11}$}
   \end{subfigure}
   \begin{subfigure}{0.45\textwidth}
      \includegraphics[width=\columnwidth,draft=false]{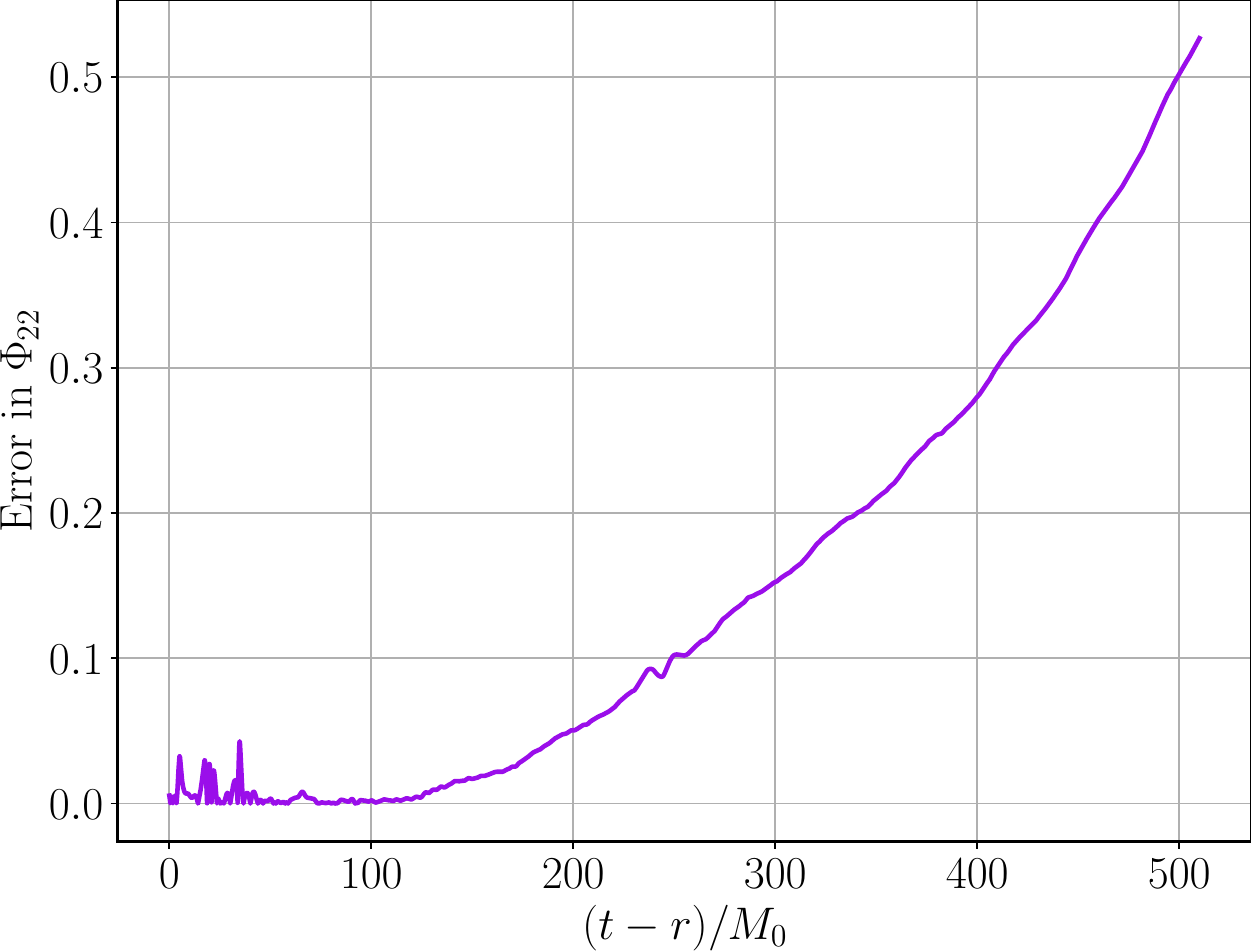}
	   \caption{Phase $\Phi_{22}$ of $\Psi_{4,22}$}
   \end{subfigure}
   \begin{subfigure}{0.45\textwidth}
      \includegraphics[width=\columnwidth,draft=false]{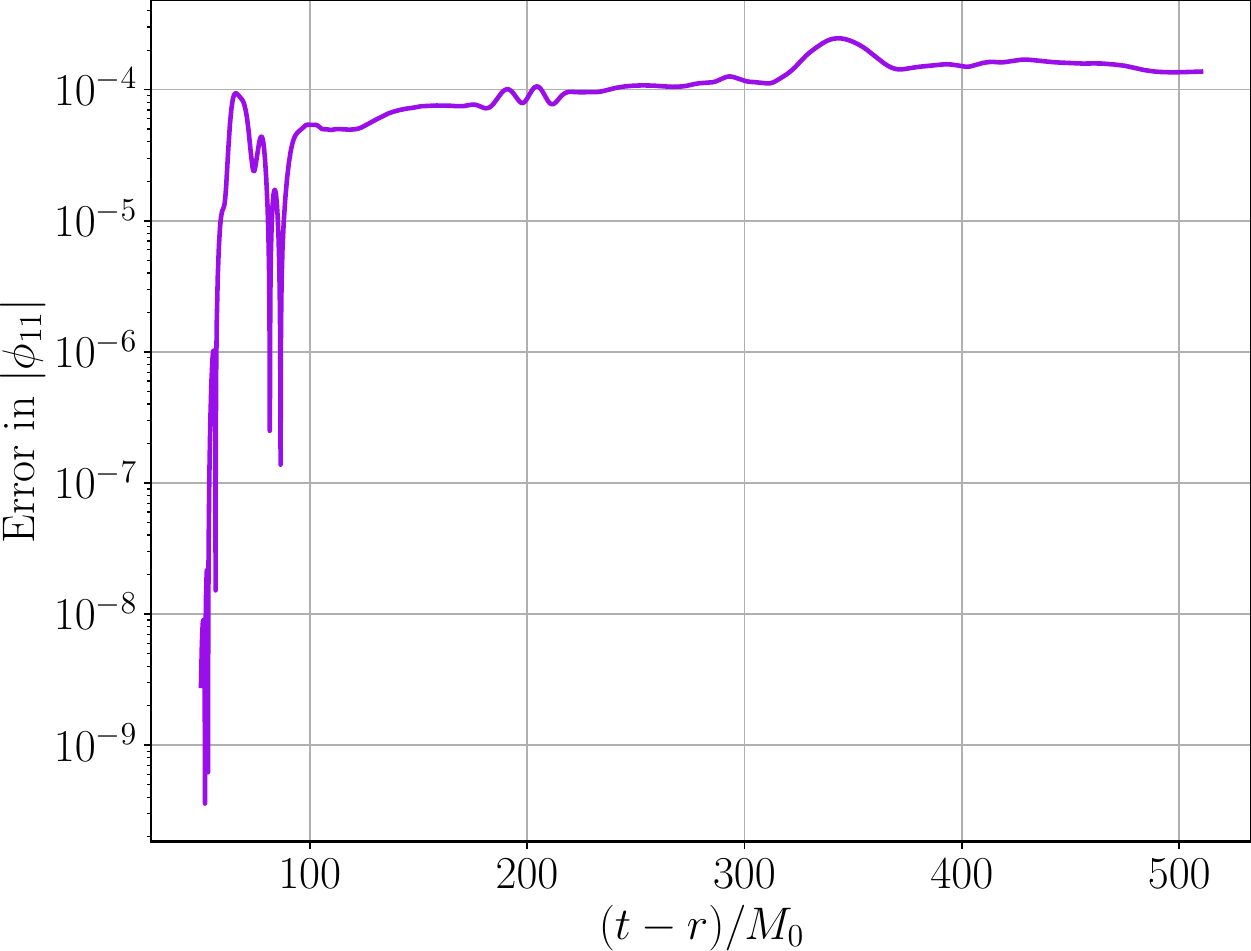}
	   \caption{Amplitude $A$ of $\phi_{11}$}
   \end{subfigure}
   \begin{subfigure}{0.45\textwidth}
      \includegraphics[width=\columnwidth,draft=false]{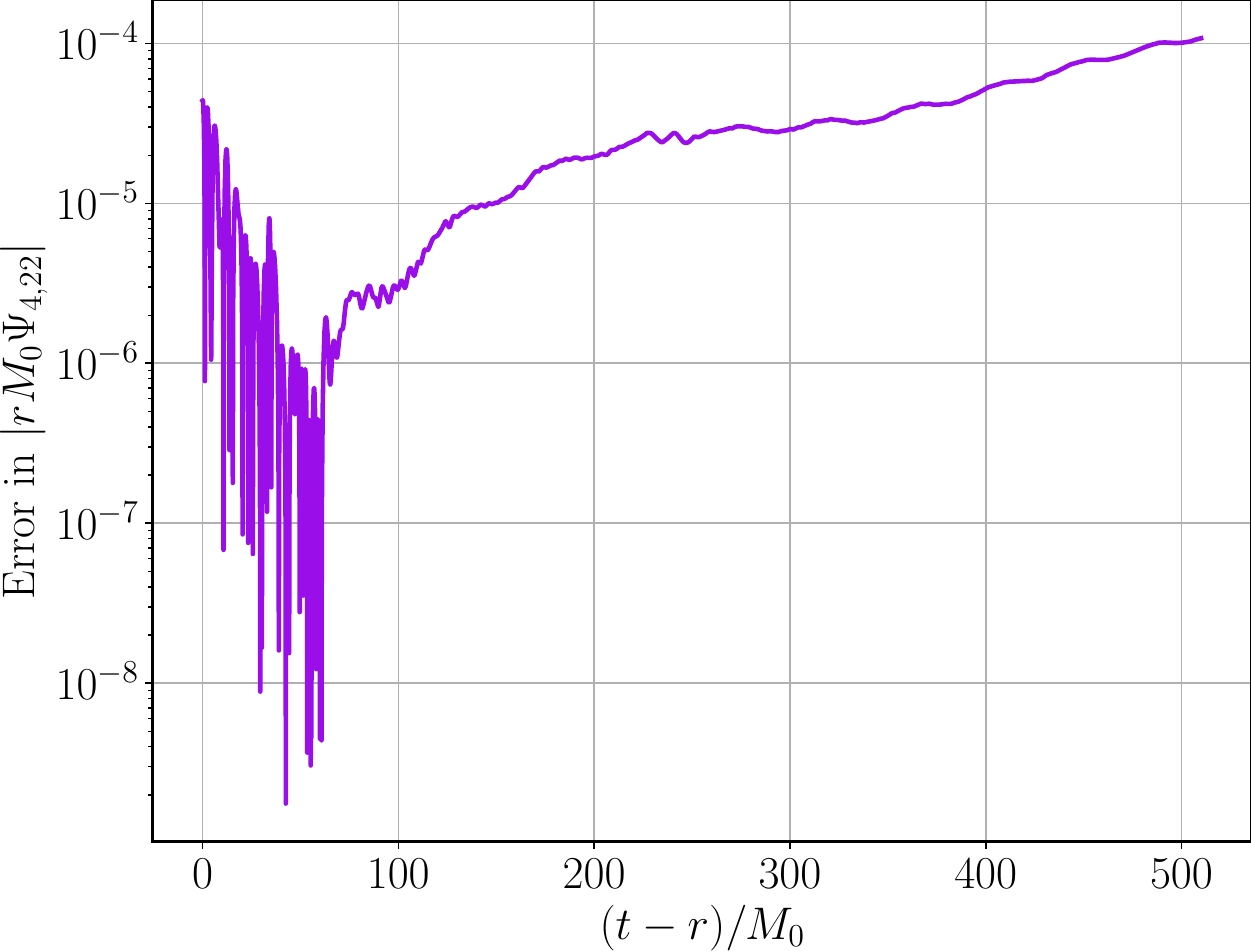}
	   \caption{Amplitude $A$ of $\Psi_{4,22}$}
   \end{subfigure}
\caption{Truncation error estimate of the medium resolution 
	obtained from the Richardson extrapolation of the   
        phase $\Phi(t)$ and amplitude $A(t)$ 
        of the scalar (left) and tensor (right) waveform extracted at $100 M_0$
	for a nonspinning BH binary with mass ratio
    $q=1/2$ and coupling $\zeta_1=0.075$.
\label{fig:RichErr_q2to1}
}
\end{figure*}

In Fig.~\ref{fig:psi4_self_conv_study}, we plot the self-convergence of the amplitude and phase
for $\Psi_{4,22}$ and $\phi_{11}$ for the $q=1/2$, $\zeta_1=0.075$ run. 
Unlike the integrated constraint violation,
we find that $\Psi_{4,22}$ and $\phi_{11}$ converge at roughly fourth order for $q=1/2$.
For the same run, we show the Richardson extrapolated error in the phase and amplitude
for $\Psi_{4,22}$ and $\phi_{11}$ (Fig.~\ref{fig:RichErr_q2to1}).

\begin{figure*}
\centering
   \begin{subfigure}{0.45\textwidth}
      \includegraphics[width=\columnwidth,draft=false]{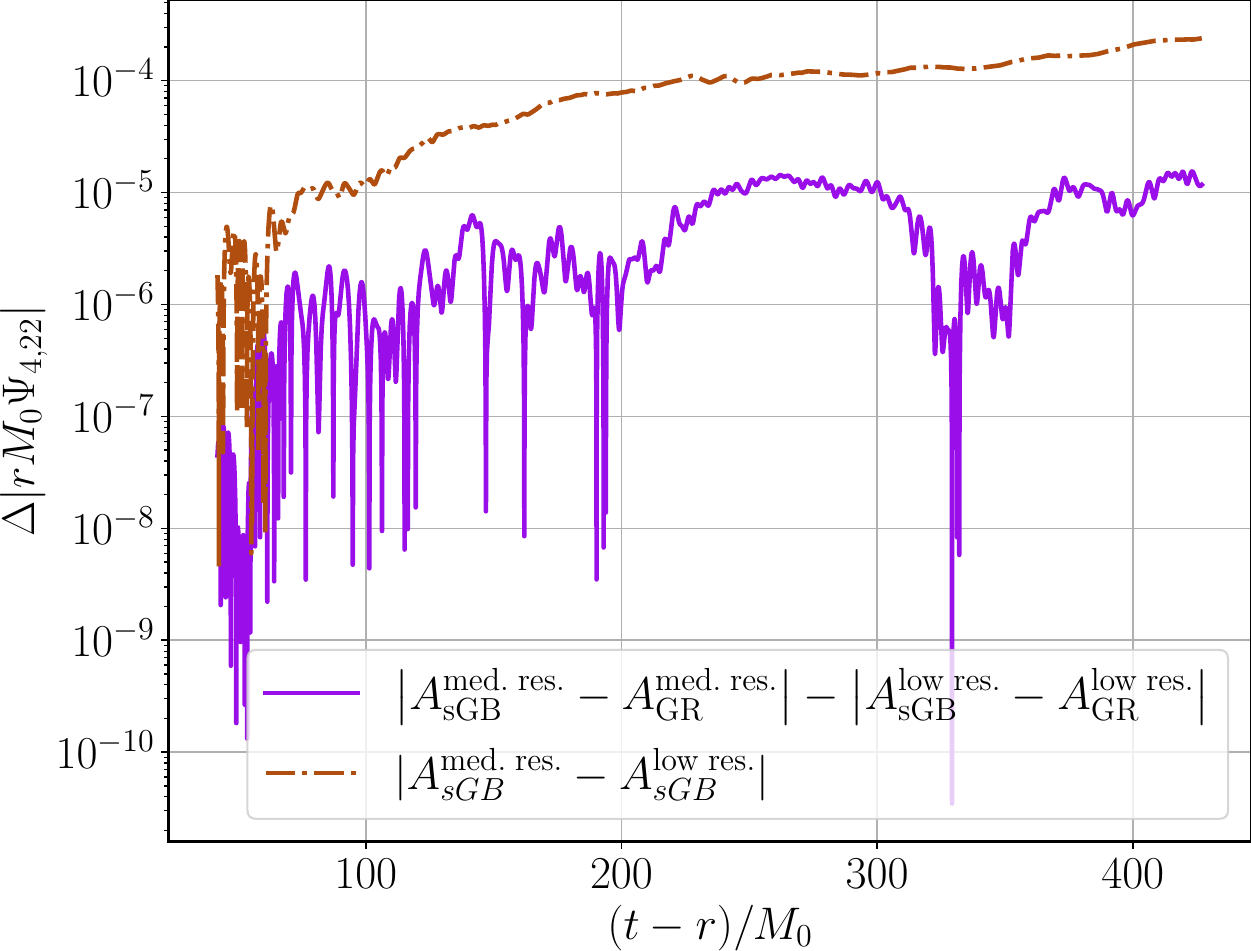}
	   \caption{Amplitude $A$ of $\Psi_{4,22}$ , $q=1/2$}
   \end{subfigure}
   \begin{subfigure}{0.45\textwidth}
      \includegraphics[width=\columnwidth,draft=false]{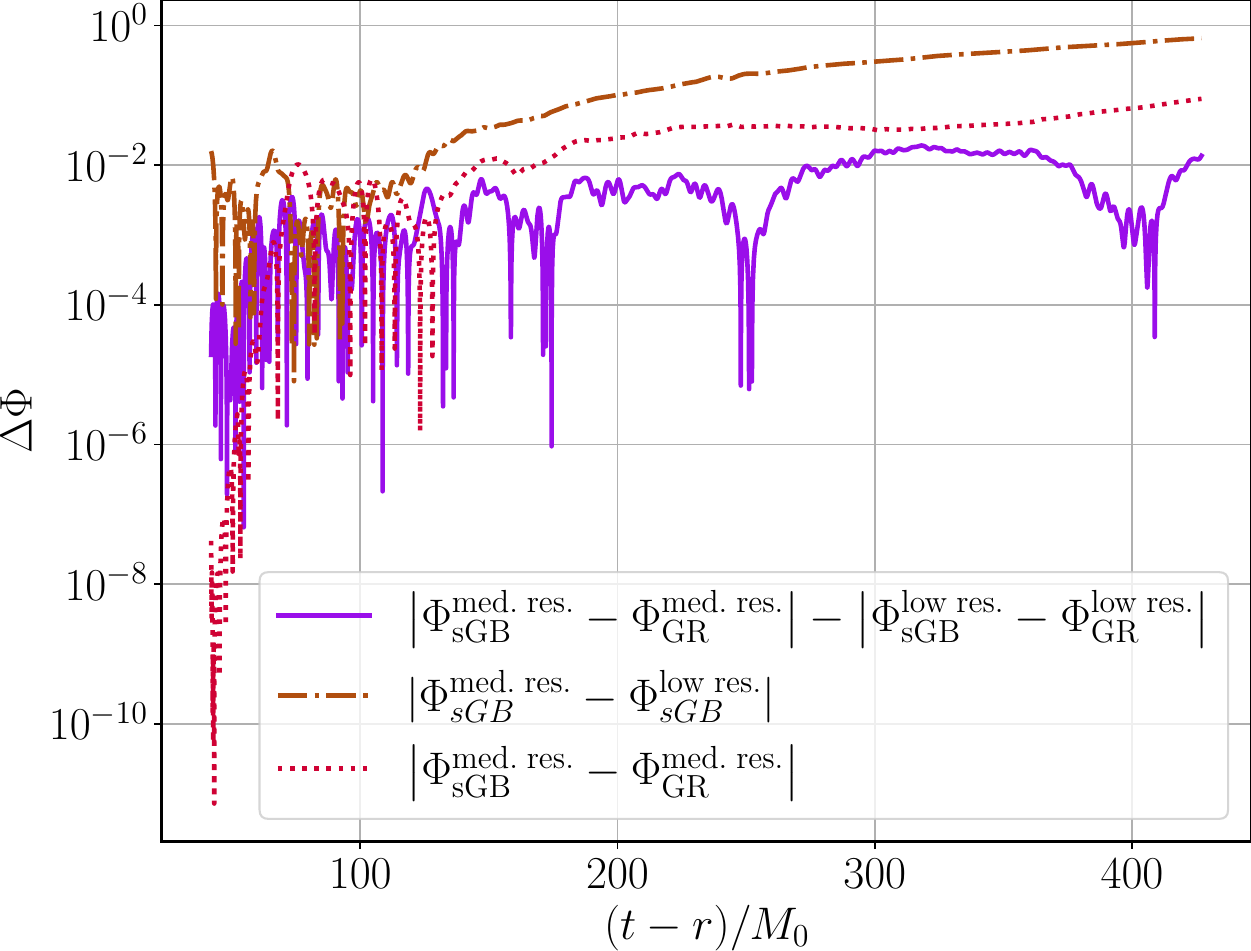}
	   \caption{Orbital phase $\Phi$ of $\Psi_{4,22}$, $q=1/2$}
   \end{subfigure}
\caption{
     We show the difference between the low and medium resolutions
     of the amplitude (left) and  phase (right) of the gravitational waveform
     for a nonspinning BH binary with mass ratio
     $q=1/2$ and coupling $\zeta_1=0.075$ (solid purple)
     and the difference of the difference between the sGB and GR amplitude and phase
     at low and medium resolutions (dashed brown line).
     This provides evidence that the truncation error roughly cancels between
     the sGB and GR runs.
\label{fig:conv_study_dephasing}
}
\end{figure*}

As discussed in Sec.~\ref{sec:results}, because we use the same numerical resolution
for carrying out the GR and sGB simulations, which we then compare to compute the dephasing
$\delta \Phi$, there is a cancellation which leads to a smaller truncation error in this
quantity compared to the overall truncation error in $\Phi$.
This is illustrated in Fig.~\ref{fig:conv_study_dephasing}, 
where we estimate the truncation error in $\delta \Phi$ by comparing
a $q=1/2$ GR simulation to an equivalent sGB simulation with $\zeta_1=0.075$ at two different resolutions.
We compare this to an estimate of the overall truncation error in $\Phi$ for the same sGB case, 
and carry out a similar comparison for the GW amplitude.

%------------------------------------------------------------------------------
\subsection{Extraction error of waveforms}

We next consider the extraction error, that is, the
errors in our waveforms due to extracting them at a finite radius.
To estimate the extraction error we compute the complex amplitude and phase of 
the $(\ell=2,m=2)$ multipole of $\Psi_4$ defined in Eq.~\eqref{eq:psi4_22} 
and the $(\ell=1,m=1)$ multipole of $\phi$ defined in Eq.~\eqref{eq:phi_decomposition}
at several extraction radii, 
and extrapolate the quantities to infinity by fitting them to polynomials in $1/r$
\begin{subequations}
\label{eq:extapol_radius}
\begin{align}
   A(r,t_{\rm ret})
   &=
	\sum\limits_{n=0}^{N} \frac{A^{(n,N)}(t_{\rm ret})}{r^n}, \\ 
   \chi(r,t_{\rm ret})
   &=
	\sum\limits_{n=0}^{N} \frac{\chi^{(n,N)}(t_{\rm ret})}{r^n}
   .
\end{align}
\end{subequations}
where $t_{\rm ret} = t - r $ refers to the retarded time,
$A$ is the amplitude of the waveform, and $\chi$ is the phase. 
The time-dependent $n=0$ coefficients are then used as the amplitude and phase of 
the asymptotic waveform.
The error from computing a field quantity 
$u (t_{\rm ret},r)$ at a finite radius $r_i$ is then 
\begin{align}
   \label{eq:extrapolation_error_formula}
   \epsilon(u,r_i,N)
   =
	| u(t_{\rm ret},r_i) - u^{(0,N)}(t_{\rm ret}) |
   .
\end{align}

\begin{figure*}
\centering
   \begin{subfigure}{0.45\textwidth}
      \includegraphics[width=\columnwidth,draft=false]{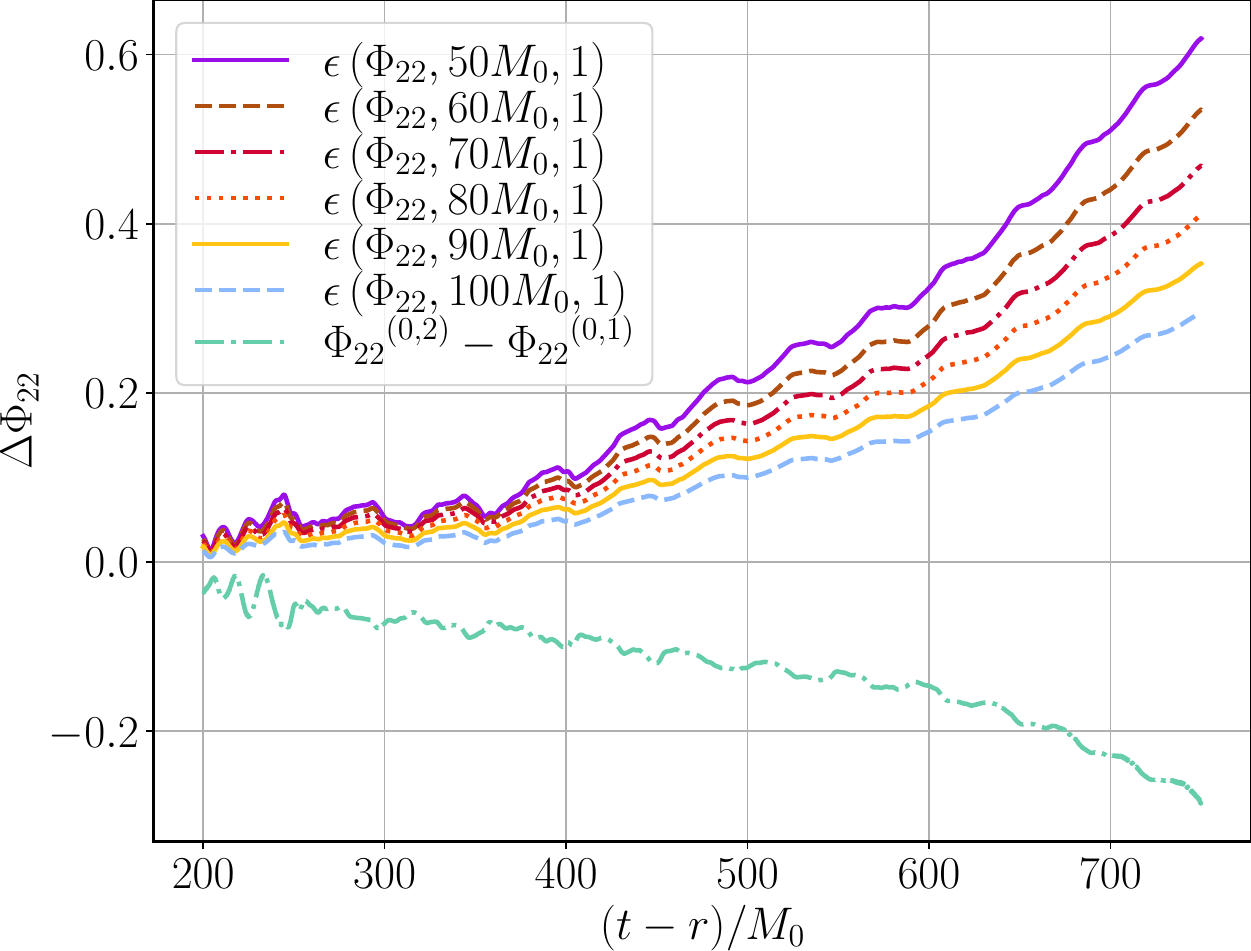}
	   \caption{Phase $\Phi_{22}$}
   \end{subfigure}
   \begin{subfigure}{0.45\textwidth}
      \includegraphics[width=\columnwidth,draft=false]{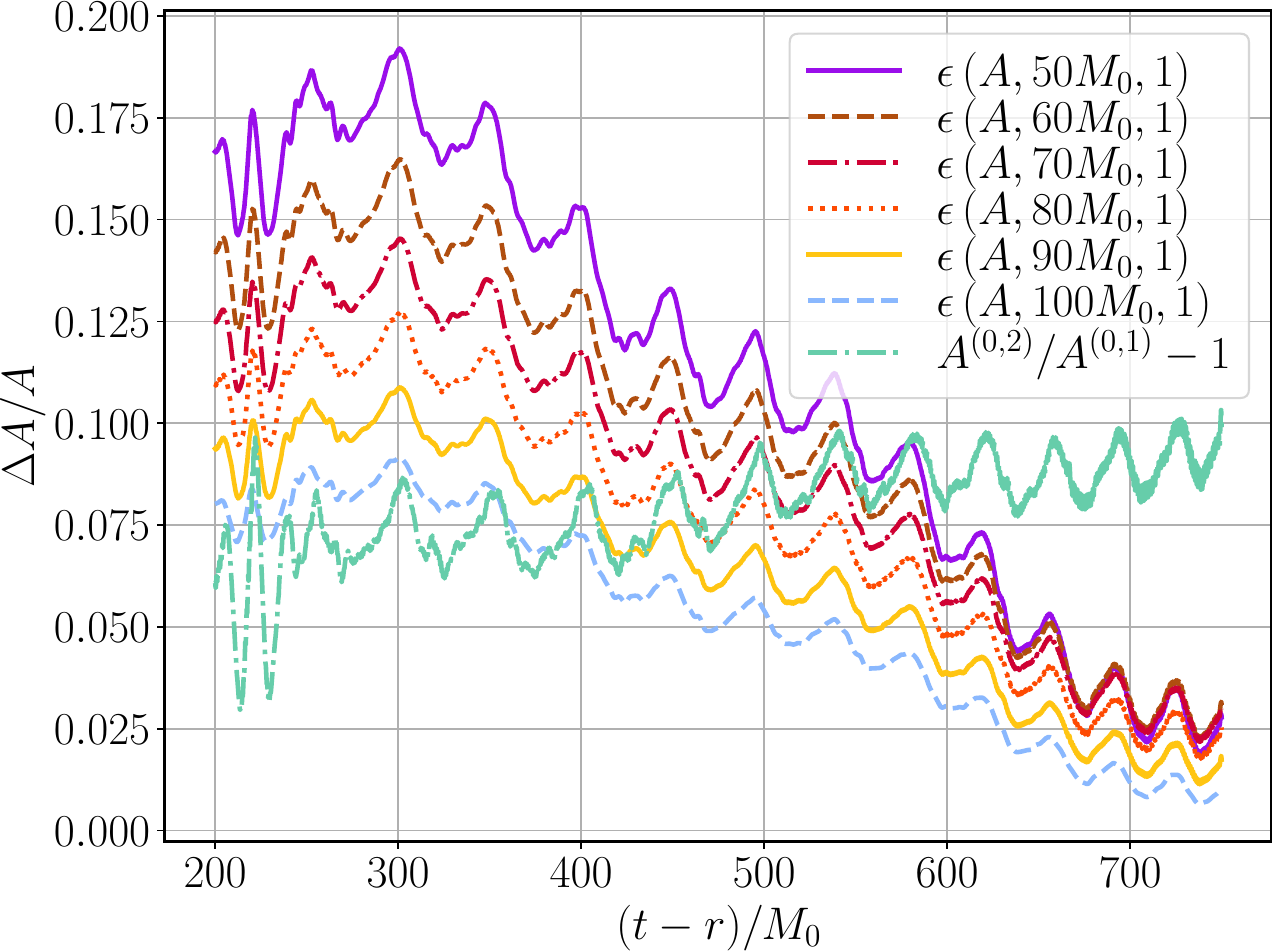}
      \caption{Amplitude $A$}
   \end{subfigure}
\caption{
	Deviation of the phase, $\epsilon(\Phi_{22},r_i,1)$ (left) and 
	relative deviation of the amplitude, $\epsilon(A,r_i,1)/A_{0,1}$ (right) of
	the waveform $r \Psi_{4,22}(t,r) M_0 =A(t,r)e^{i \Phi_{22}(t,r)} $ obtained
     at finite extraction radius from the values extrapolated according to 
     Eq.~\eqref{eq:extapol_radius} for a nonspinning BH binary with mass ratio
     $q=1/2$ and coupling $\zeta_1=0.075$.
\label{fig:extraction_error}
}
\end{figure*}

\begin{figure*}
\centering
   \begin{subfigure}{0.45\textwidth}
      \includegraphics[width=\columnwidth,draft=false]{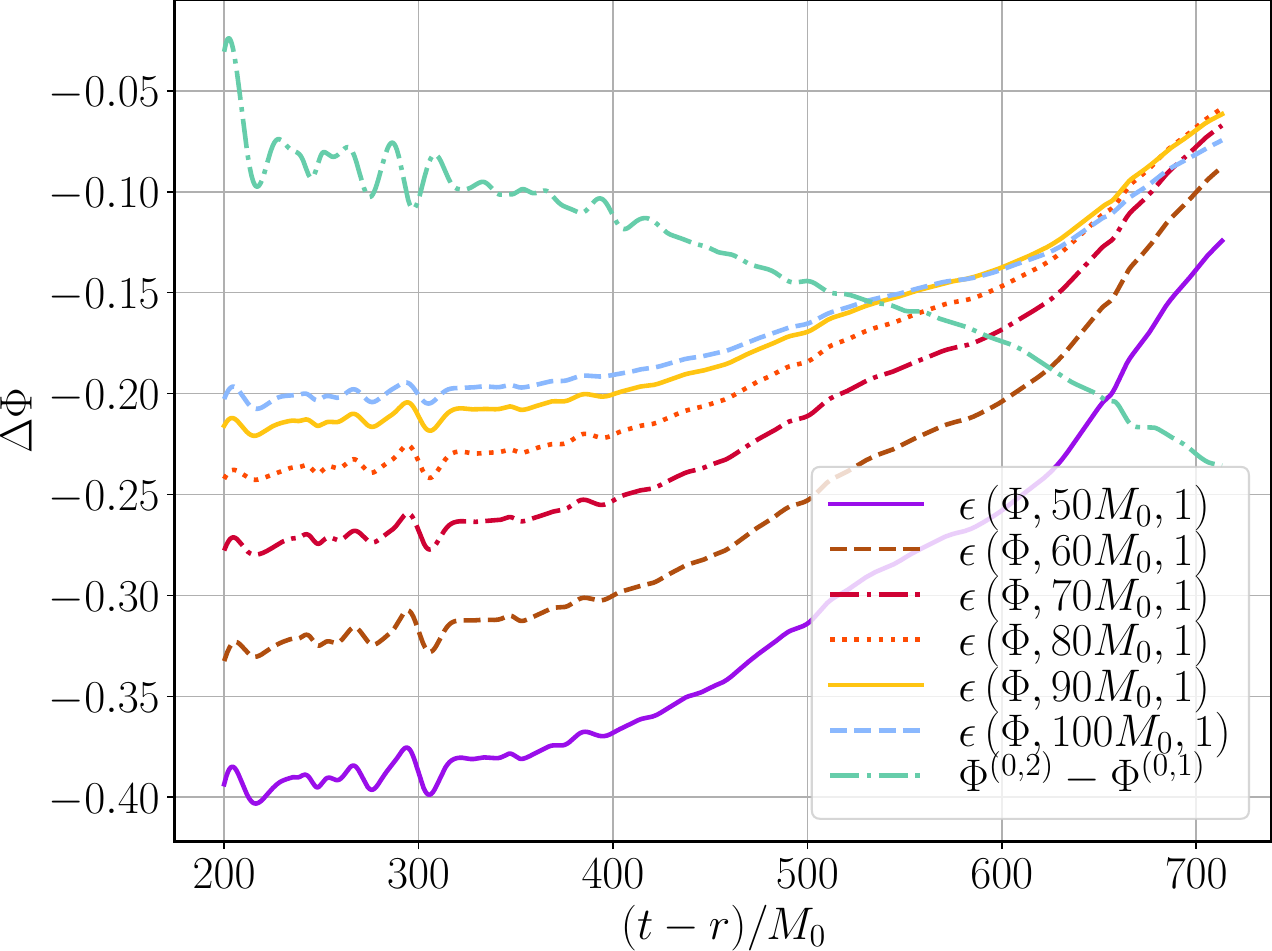}
      \caption{Phase $\Phi$}
   \end{subfigure}
   \begin{subfigure}{0.45\textwidth}
      \includegraphics[width=\columnwidth,draft=false]{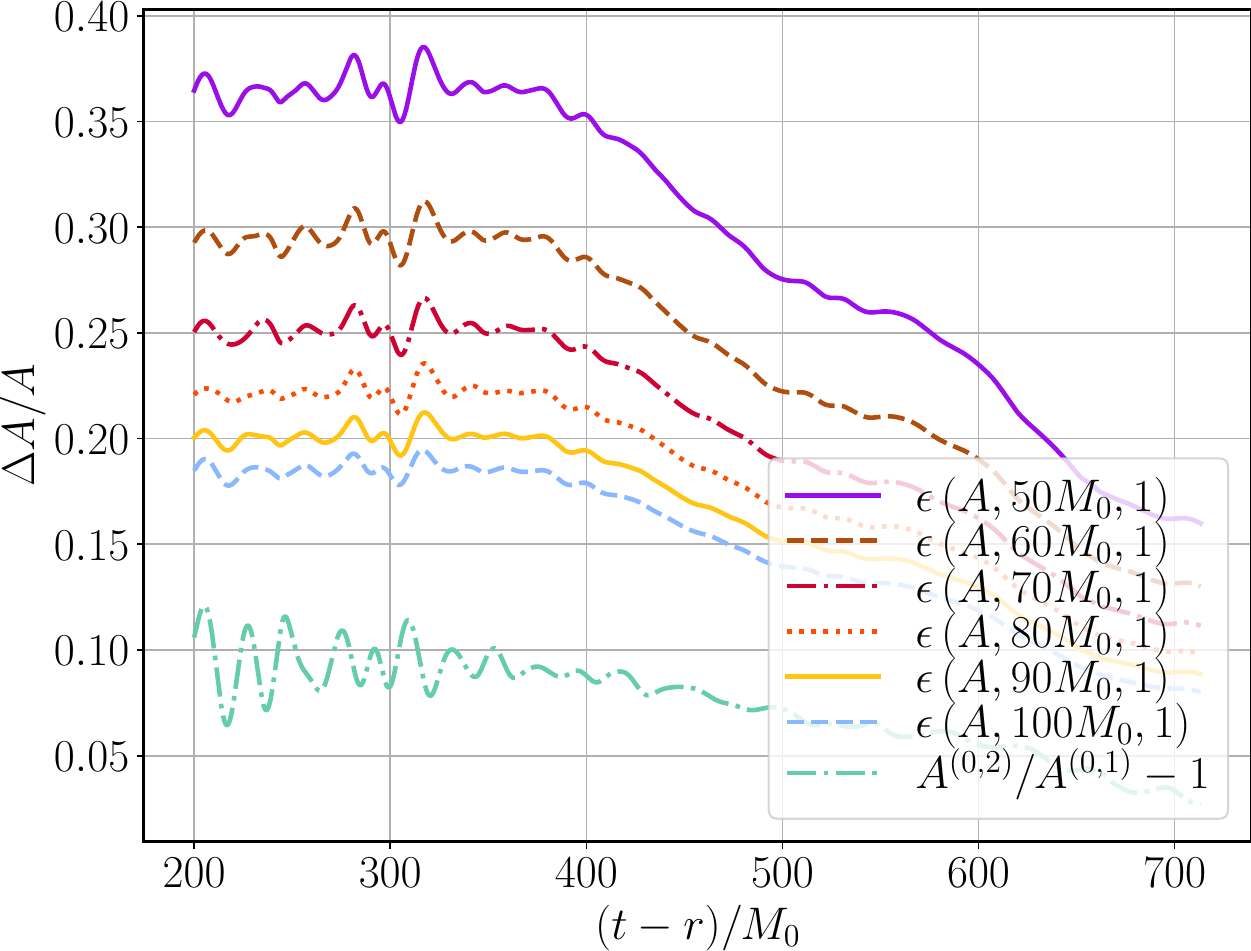}
      \caption{Amplitude $A$}
   \end{subfigure}
\caption{
     Deviation of the phase, $\epsilon(\Phi,r_i,1)$ (left) and relative deviation of 
	the amplitude, $\epsilon(A,r_i,1)/A_{0,1}$  (right) of
	the waveform $(r/M_0) \phi_{11}(t,r)  =A(t,r)e^{i \Phi(r,t)} $ obtained
     at finite extraction radius from the values extrapolated according to 
     Eq.~\eqref{eq:extapol_radius} for a nonspinning BH binary with mass ratio
     $q=1/2$ and coupling $\zeta_1=0.075$.
\label{fig:extraction_error_sf}
}
\end{figure*}

In Fig.~\ref{fig:extraction_error} and~\ref{fig:extraction_error_sf}, 
we plot our estimates for the error
due to the extraction of the gravitational and scalar waveforms
at a finite radius. 
Comparing these to the estimate of the 
truncation error in Fig.~\ref{fig:RichErr_q2to1}
we conclude that the finite resolution of the code is the dominant source
of error.

\subsection{Orbital eccentricity}
To estimate the orbital eccentricity of the binary system, 
introduced by imperfect initial data, 
we use the gravitational wave phase \cite{Mroue:2010re}. 
We write the $\left(\ell,m\right)=\left(2,2\right)$ component of $\Psi_4$
in the wave zone as:
\begin{align}
   \label{eq:psi4_22}
   r \ M_0 \times \Psi_{4,22}
   \equiv
   A_{22}(t,r)
   e^{-i\Phi_{22}}
   +
   \mathcal{O}\left(\frac{1}{r}\right)
   .
\end{align}
We fit a $5^{\rm th}$ order polynomial to the orbit-averaged $\Phi$ to obtain
$\Phi_{\rm fit}$, and define the eccentricity to be the amplitude of 
the oscillating function
\begin{align}
   \label{eq:eccentricity}
   e_{\Phi}(t)
   \equiv
   \frac{\Phi_{22} (t)-\Phi_{\rm{fit},22}(t)}{4}
   .
\end{align}

\begin{figure*}
\centering
   \begin{subfigure}{0.45\textwidth}
      \centering
      \includegraphics[width=\columnwidth,draft=false]{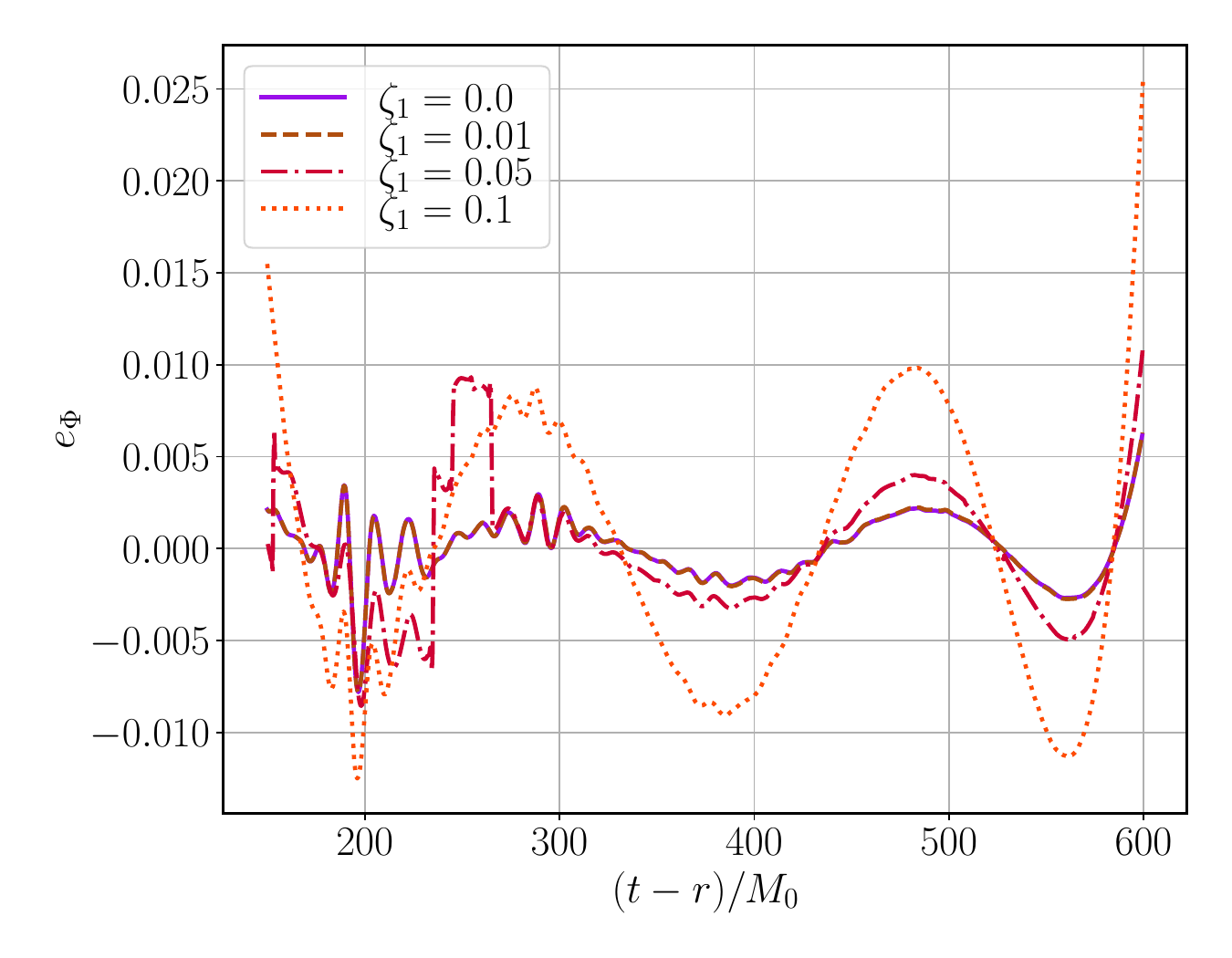}
      \caption{$q=1$}
      \label{fig:eccentricity_1t1}
   \end{subfigure}
   \begin{subfigure}{0.45\textwidth}
      \centering
      \includegraphics[width=\columnwidth,draft=false]{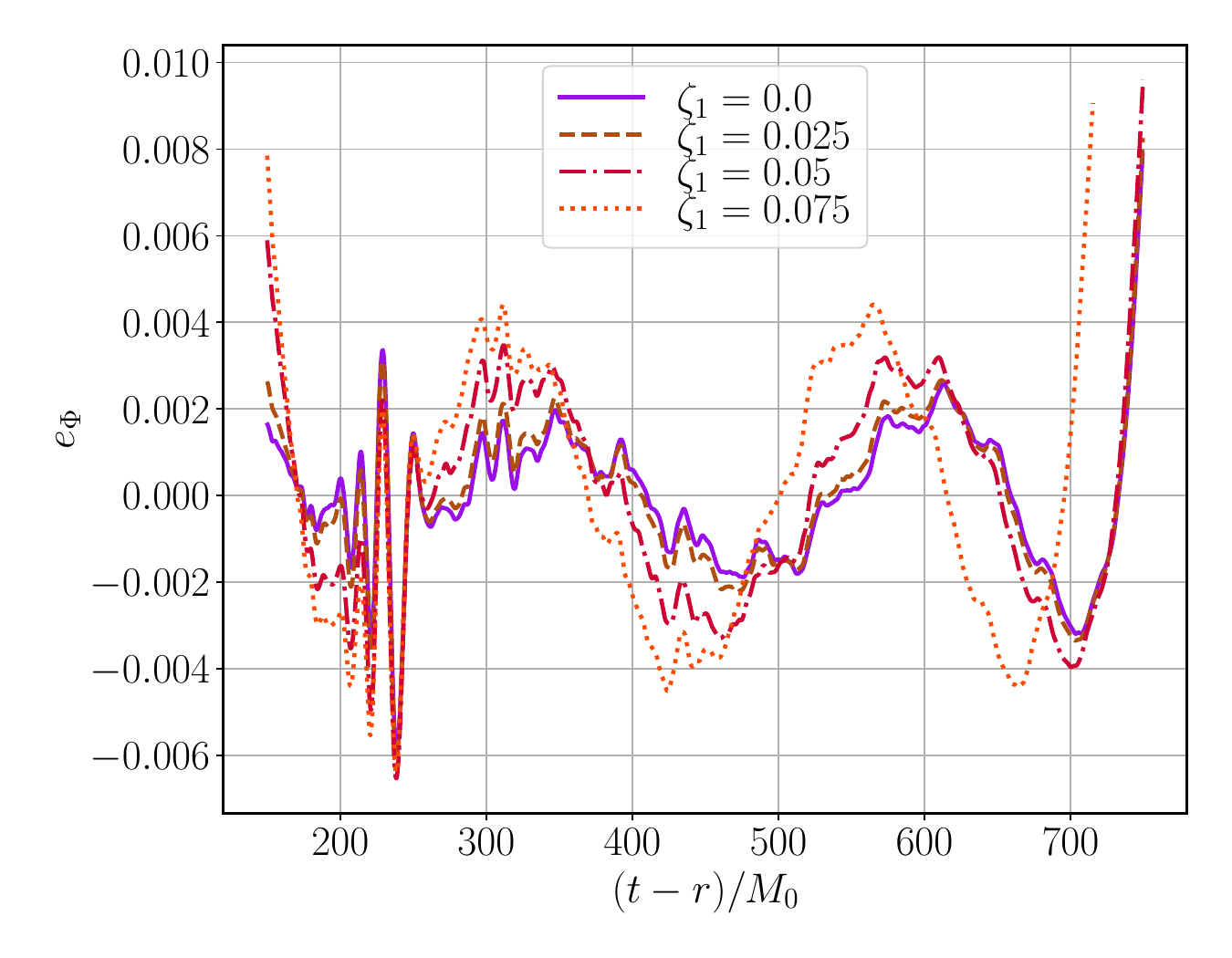}
      \caption{$q=2/3$}
      \label{fig:eccentricity_3t2}
   \end{subfigure}
   \begin{subfigure}{0.45\textwidth}
      \centering
      \includegraphics[width=\columnwidth,draft=false]{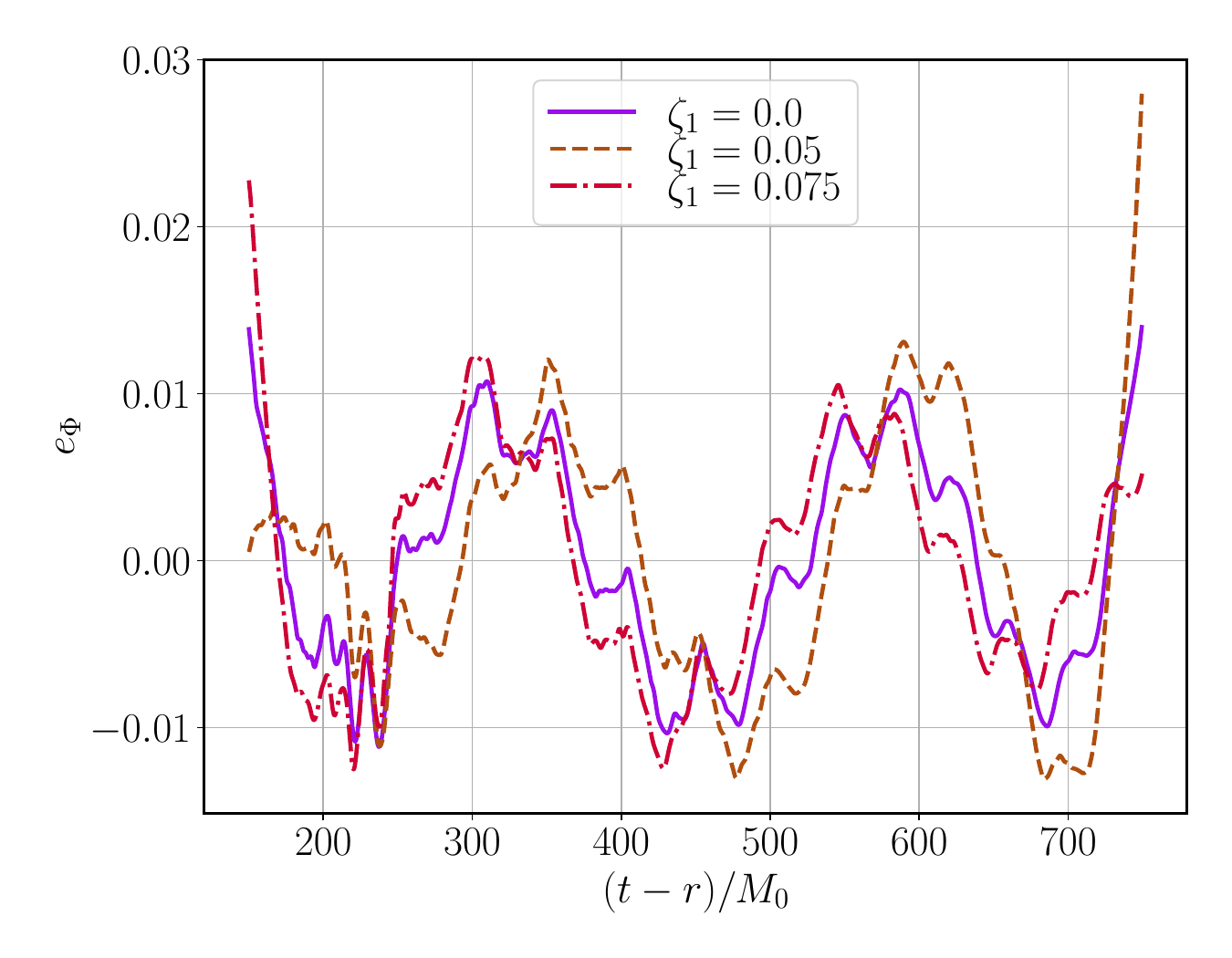}
      \caption{$q=1/2$}
      \label{fig:eccentricity_1t2}
   \end{subfigure}
   \begin{subfigure}{0.45\textwidth}
      \centering
      \includegraphics[width=\columnwidth,draft=false]{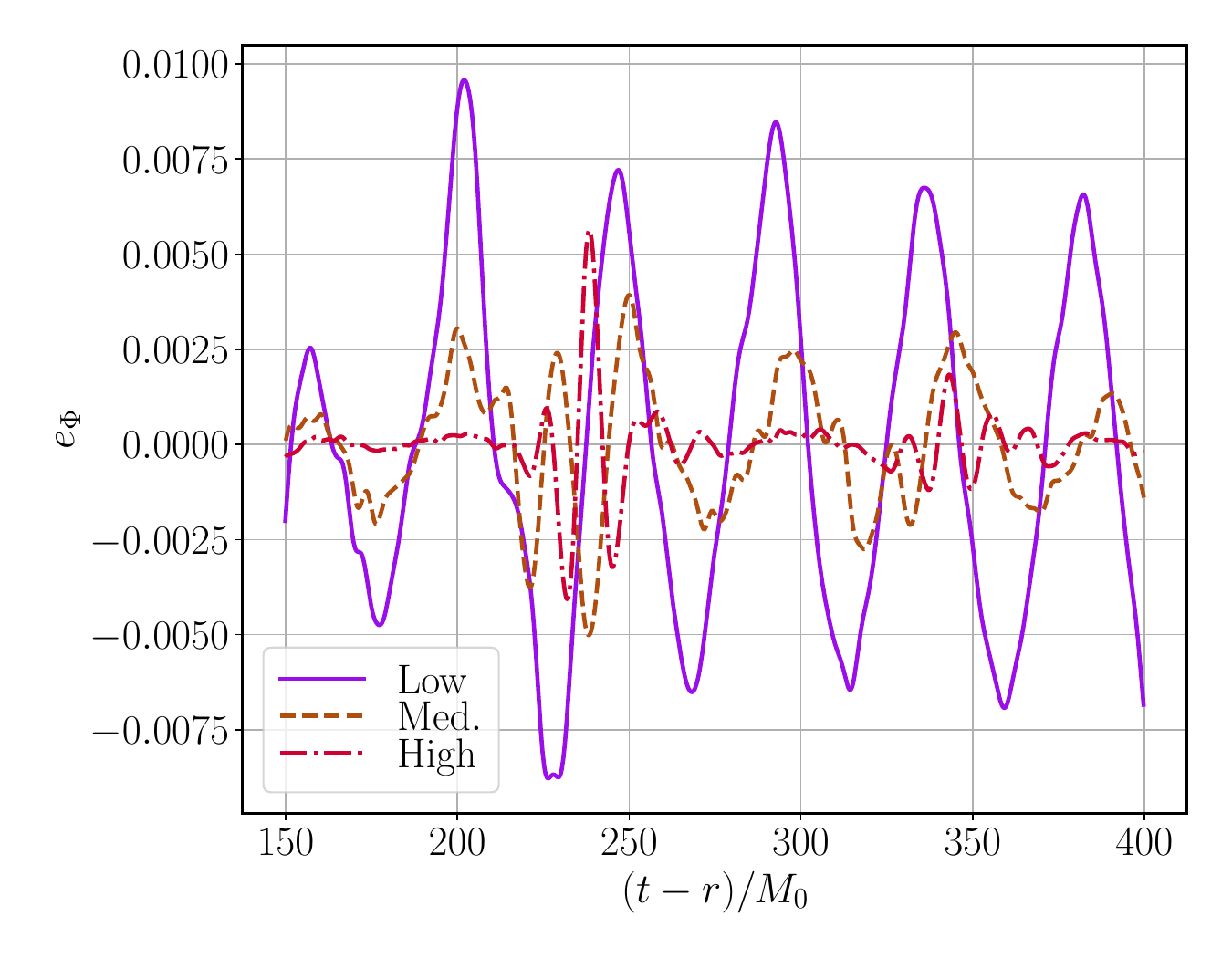}
      \caption{$q=2/3, \zeta_1=0.075$, resolution study}
      \label{fig:eccentricity_3t2_res}
   \end{subfigure}
\caption{
   Eccentricity estimator of the $q=1$, $q=2/3$, and $q=1/2$ mass ratio
   inspirals for several different values of $\zeta_1$, and
   for the $q=2/3$ mass ratio at different resolutions.
   We see that the eccentricity of the binaries we study is affected by both
   the form of our initial data (the formation of scalar charge from vacuum 
   initial conditions), and, to a greater extent, from the resolution of our runs.
	We measure the eccentricity using the radially extrapolated Weyl scalar $\Psi_{4,22}$.
	The low and high resolution have $2/3$ and $4/3\times$ the resolution
	of the medium resolution which has a linear grid spacing of $dx = 0.006M_0$
	on finest level.
\label{fig:eccentricity}
}
\end{figure*}

We plot the eccentricity [see Eq.~\eqref{eq:eccentricity}] of our simulations
in Fig.~\ref{fig:eccentricity} for different values of $\zeta_1$ and
resolution.  Ideally, an eccentricity estimator will plot a sinusoidal wave as
a function of time \cite{Mroue:2010re}. Our eccentricity measurements have
higher harmonics, which we attribute to the junk radiation from the choice of
puncture initial data, and from the black hole scalarization process, and from
the fact that we only measure the eccentricity over a relatively short inspiral
time ($t/M_0<1000$).  While the eccentricity does slightly increase with
increasing $\zeta_1$, we find that our eccentricity is mostly limited by
resolution, and not from perturbations caused by our initial data.  This
suggests that the dephasing between the sGB and GR simulations is not dominated
by small differences in the eccentricity of our simulations  caused by the
rapid development of the scalar field around the black holes at early times.

%=============================================================================
\section{Post-Newtonian results in sGB gravity\label{sec:pn_theory}}
   Due to the presence of monopole scalar charge around each black hole
in sGB gravity, black hole inspirals can emit scalar radiation,
which enters at $-1$PN order as dipole emission
for unequal mass black hole binaries. 
The calculation of the leading PN 
correction to the gravitational and scalar
radiation for binary black holes in sGB gravity was carried out
in Ref.~\cite{Yagi:2011xp}.
In the limit of an exactly equal mass, nonspinning binary, the dipole
radiation vanishes. More generally it is
straightforward to see that any odd multipole of a scalar
is zero in this case as the spherical harmonics are odd under
parity inversion ($\vec{r}\to-\vec{r}$), 
but the spacetime in this case is even under this transformation.
Thus, for equal mass black hole binaries, the scalar waveform
enters at higher PN order 
\cite{Yagi:2011xp,Shiralilou:2020gah,Shiralilou:2021mfl}.

The PN calculations initiated in Ref.~\cite{Yagi:2011xp} were
recently extended to higher PN order in 
Refs.~\cite{Shiralilou:2020gah,Shiralilou:2021mfl}.
In those works, the authors additionally considered more 
general Gauss-Bonnet couplings $f(\phi)\mathcal{G}$. 
Here we only present the leading-order PN results.
To leading order in $\zeta_1$, spherical harmonic components
of the scalar radiation of the binary system go as 
~\cite{Yagi:2011xp,Witek:2018dmd,Shiralilou:2020gah,Shiralilou:2021mfl})
\begin{subequations}
\label{eq:pn_formulas_integrated_scalar_field}
\begin{align}
   \phi_{00}
   &\approx
   \left(\frac{2\lambda}{r}\right)
   \left(8\pi\right)^{1/2}
   \frac{M_0}{m_1m_2}
   ,\\
   \phi_{11}
   &\approx
   -
   \left(\frac{2\lambda}{r}\right)\left(\frac{2\pi}{3}\right)^{1/2} 
   \left(
      1
      +
      \frac{3m_1^2 + 3m_2^2 + 4m_1m_2}{M_0^2}
      x
   \right)
   \frac{\Delta M_0}{m_1m_2}x^{1/2}
   ,\\
   \phi_{22}
   &\approx
   - \left(\frac{2\lambda}{r}\right) 
   \left(\frac{8\pi}{15}\right)^{1/2}
   \frac{m_1^2 - m_1m_2 + m_2^2}{M_0m_1m_2}x
   ,\\
   \phi_{33}
   &\approx
   \left(\frac{2\lambda}{r}\right) 
   \left(\frac{1296\pi}{35}\right)^{1/2}\frac{\Delta M\left(m_1^2+m_2^2\right)}{8M_0^2m_1m_2}
   x^{3/2}
   ,\\
   \phi_{44}
   &\approx
   \left(\frac{2\lambda}{r}\right) 
   \left(\frac{2048\pi}{315}\right)^{1/2}
   \frac{
      m_1^4
      -
      m_1^3m_2
      +
      m_1^2m_2^2
      -
      m_1m_2^3
      +
      m_2^4
   }{
      3m_1m_2M_0^3
   }x^2
   ,
\end{align}
\end{subequations}
where $m_{1,2}$ are the masses of the two black holes, 
with the convention $m_1\leq m_2$ (see Sec.~\ref{sec:initial_data}), 
$\Omega$  
is the angular velocity of the binary in the center of mass frame, and
\begin{subequations}
\begin{align}
   \phi_{\ell m}
   &\equiv
   \lim_{r\to\infty}
   \int_{\mathbb{S}_2} 
   \bar{Y}_{lm}
   \phi 
   ,\\
   M_0
   &\equiv
   m_1
   +
   m_2
   ,\\
   \Delta M
   &\equiv
   m_2
   -
   m_1
   ,\\
   \label{eq:definition_x0}
   x
   &\equiv
   \left(M_0\Omega\right)^{2/3}
   .
\end{align}
\end{subequations}
Note that the second terms in 
Eq.~\eqref{eq:pn_formulas_integrated_scalar_field}, which are raised to the $1/2$ power,  come from the
integral over the sphere of $\bar{Y}_{lm}$.
The scalar waveforms 
Eq.~\eqref{eq:pn_formulas_integrated_scalar_field}
are presented to leading order in the PN
expansion, except for the $\ell=m=1$ waveform, which has been
computed to $0.5$PN order \cite{Shiralilou:2020gah,Shiralilou:2021mfl}.

We next consider the dephasing of gravitational waves in PN theory.
We write the orbital phase in the time domain as a function of the PN parameter $x$,
\begin{align}\label{eq:def_dephasing}
   \Phi(x)
   =
   \Phi_{\rm GR}(x)
   +
   \delta\Phi(x)
   .
\end{align}
Here, $\Phi_{\rm GR}$ is the orbital phase when setting $\lambda=0$, and $\delta\Phi$
is the additional phase shift that comes from the emission of scalar radiation.
In the PN expansion of scalarized compact objects, 
there are two limits considered in the literature: 
the \emph{dipole driven regime} and the \emph{quadrupole driven regime}
\cite{Sennett:2016klh,Shiralilou:2021mfl}. 
In the dipole driven regime, the dipole scalar emission is the dominant
source of radiated energy, while in the quadrupole driven regime, the
dominant source of radiated energy is the gravitational wave emission.
The system is in the quadrupole driven regime when 
\begin{equation}
   x \gtrsim \frac{5}{24} {\mathcal{S}_{-}}^2 
\end{equation}
where we introduced the scalar dipole
\begin{equation}
   \mathcal{S}_{\pm} 
   \equiv 
   \frac{{\alpha}_2 \pm {\alpha}_1}{2 \sqrt{\bar{\alpha}}} ,
\end{equation}
where $\bar{\alpha} \equiv \left(1 +{\alpha}_1 {\alpha}_2\right)$,
and $\alpha_i$ are the \emph{black hole sensitivities}
for sGB gravity \cite{Julie:2019sab,Julie:2022huo} (for their
explicit values, see Eq.~\eqref{eq:bh_gb_sensitivity} below).
Notice, for equal mass ratio binaries, the system is always in the quadrupole
driven regime as there is no dipole radiation ($\mathcal{S}_-=0$).
We see that the system is in the dipole driven regime only for 
unequal mass ratio binaries that are far apart (that is, when $x$ is small).
Given the experimental constraints on $\zeta_1\ll1$ and $\mathcal{S}_{-}$,
the binary systems of interest for ground- and space-based GW detectors
are driven by the quadrupolar driven regime for sGB gravity.
We thus compare our numerical waveforms to gravitational waveforms for systems
in which quadrupolar radiation is dominant. 

The leading order contribution to the GW phase
in ESGB gravity was computed in
Refs.~\cite{Yagi:2011xp,Yagi:2012gp} using the stationary-phase-approximation \cite{Maggiore:2007ulw}, 
and later extended to higher orders in PN theory in 
Refs.~\cite{Shiralilou:2020gah,Shiralilou:2021mfl,Lyu:2022gdr}.
The highest order PN corrections to the phase so far have been computed
by Lyu et al. \cite{Lyu:2022gdr}, 
who mapped results obtained partially to 2PN order 
in scalar-tensor theories \cite{Sennett:2016klh} to sGB gravity.
Here, we review their calculation, and present results for the 
time-domain orbital
phase $\delta \Phi$ as a function of the PN parameter $x$.

The results of Ref.~\cite{Sennett:2016klh} were presented in the Jordan frame,
and ESGB gravity is written in the Einstein frame. Thus, the first step
Lyu et al. took was to transform the results of Ref.~\cite{Sennett:2016klh} to the
Einstein frame. After this transformation, Lyu et al. noticed that the
results of Ref.~\cite{Sennett:2016klh} were expressed in terms of the
black hole sensitivities 
$\alpha_i$, and their derivatives $\beta_i$. 
These were computed for black holes in ESGB gravity 
by Juli\'{e} et al.~\cite{Julie:2019sab,Julie:2022huo},
and for non-spinning black holes are given by (here we used the conversion
$\upvarphi \to \phi/\sqrt{2}$, $f(\upvarphi) \to 2\sqrt{16\pi} \upvarphi$, and 
$\alpha_{\rm GB} \to \lambda/\sqrt{8 \pi}$)
\begin{align}
   \label{eq:bh_gb_sensitivity}
   \alpha_i 
   &\equiv 
   -
   \frac{\alpha_{\rm GB} f'(\upvarphi_0)}{2 {m_i}^2} 
   = 
   - \frac{\sqrt{2}\lambda}{{m_i}^2} 
   ,
   \\
   \beta_i 
   &\equiv 
	\frac{d \alpha_i}{d \upvarphi}\rvert_{\upvarphi_0} 
   = 
   -
   \frac{{\alpha_{\rm GB}}^2 f'(\upvarphi_0)^2}{2 {m_i}^2} 
   = 
   - 
   \frac{4 \lambda^2}{{m_i}^2}
   ,
\end{align}
where $\upvarphi_0 $ is the asymptotic value of scalar field at infinity 
(we set $\upvarphi_0=0$).
We see that $\beta_i\propto\lambda^2$, so it is negligible compared to $\alpha_i$.
Using these expressions, and keeping terms up to $\mathcal{O}(\lambda^2)$, 
sGB corrections to the orbital phase in the 
quadrupolar driven regime can be expressed as
\begin{equation}\label{eq:dphi_pn}
   \delta \Phi (x) 
   = 
   \sum_i \delta \Phi_{i,\rm PN} 
   = 
   \frac{\lambda^2}{8\pi m_1^4 m_2^4 \eta} \sum_i c_i x^{(-5+2 i)/2}
   ,
\end{equation}
where
\begin{eqnarray}\label{eq:pn_coeff}
   c_{-1} 
   &=& 
   \frac{25 \pi}{1344} \left(m_2^2-m_1^2\right)^2
   \\
   c_{0} 
   &=& 
   \frac{5\pi}{32256} \left[(659+728 \eta)(m_2^2-m_1^2)^2\right] +\frac{5\pi}{12} m_2^2 m_1^2 
   \\
	c_{0.5} &=& -\frac{25 \pi}{384}\left(m_2^2-m_1^2\right)^2 \left( 3 \pi +{f}_3^{\mathrm{ST}}\right)
   \\
   c_{1} &=&
   \frac{5 \pi}{585252864}\bigg[55883520(m_2^3 m_1 +m_2 m_1^3)
   +
   25(1640783 + 2621304 \eta + 2095632 \eta^2)(m_2^4+m_1^4) 
   \nonumber 
   \\
   && 
   -
   2 m_1^2 m_2^2 (83960375 + 43179192 \eta + 52390800 \eta^2 ) \bigg] 
   -
   \frac{25\pi}{288}\left(m_2^2-m_1^2\right)^2 {f}_4^{\mathrm ST}
   \\
   c_{1.5} 
   &=& 
   -
   \frac{5 \pi^2}{12}\left(m_2^4-14 m_1^2 m_2^2+m_1^4\right)
   -
   \frac{5 \pi}{96}\left(m_2^4-6 m_1^2 m_2^2+m_1^4\right){f}_3^{\mathrm ST}
   +
   \frac{1}{\lambda^2}\frac{5\pi}{8}m_1^4m_2^4 f_3^{\rm ST}
   \\
   c_2 
   &=& 
   \frac{5 \pi}{48771072 }\bigg[-24385536(m_2^3 m_1+m_2 m_1^3)
   +
   (4341025 - 65553264 \eta + 684432 \eta^2)(m_1^4+m_2^4) 
   \nonumber 
   \\
   &&
   +
   54 m_1^2 m_2^2 (-12500965 + 19310256 \eta + 366128 \eta^2)\bigg]
   \\
   &&
   -
   \frac{5 \pi}{48}\left(m_2^4-14 m_1^2 m_2^2+m_1^4\right){f}_4^{\mathrm ST}
   +
   \frac{1}{\lambda^2}
   \frac{5\pi}{4}
   m_1^4m_2^4 
   f_4^{\rm ST} 
   .
\end{eqnarray}
and $\eta \equiv m_1 m_2/M_0^2$ is the symmetric mass ratio.
Our calculation of these coefficients are presented in an ancillary 
Mathematica notebook.
As noted in Ref.~\cite{Lyu:2022gdr},
the leading $-1\rm PN$ term here agrees with the one found in 
Refs.~\cite{Yagi:2011xp,Yagi:2012gp}.
We note that we have \emph{not} included black hole spin dependence here
(in the notation of Ref.~\cite{Lyu:2022gdr}, we have set $s_i=1$, 
although the notebook presents results for general $s_i$.
The terms at $0.5$ PN onwards contain currently unknown 
coefficients ${f}_{2n}^{\rm ST}$, 
which represent our ignorance of the new scalar contributions at relative 
$n = 1.5$ and $ n = 2$ PN order in the non-dipolar flux 
(part of the flux that does not vanish for an equal mass binary)
beyond 1PN order \cite{Sennett:2016klh}; we see that $f_{2n}^{\rm ST}$
must scale as $\lambda^{2+n}, \; n>0$ in order for these terms to not
be important as $\lambda\to0$. 
In the quadrupolar driven regime, experimental constraints
on the weak-field parameters of scalar-tensor gravity suggest 
that these contributions should be much smaller
than the 2PN GR terms \cite{Sennett:2016klh}, so $f_{2n}^{\rm ST}$ is set to zero
in Ref.~\cite{Lyu:2022gdr} and in this work.
%=============================================================================
\section{\label{sec:puncture_id}
Puncture initial data for sGB binary black hole evolution
}
As we discuss in Sec.~\ref{sec:initial_data},
the Hamiltonian and momentum constraint equations in sGB gravity
reduce to those of GR when $\phi=\partial_t\phi=0$ on the initial data
hypersurface \cite{East:2020hgw,Ripley:2022cdh},
and we make use of GR puncture initial data in our simulations.
While puncture initial data is well known \cite{Cook:2000vr} and
the \texttt{TwoPunctures} implementation of that formalism
is widely used \cite{Ansorg:2004ds}, to our knowledge it has never
been implemented in conjunction with black hole excision and a (modified)
generalized harmonic formulation.
Here, we review puncture initial data, and how we incorporated
the \texttt{TwoPunctures} initial data in our MGH code.

First we write the metric in ADM variables:
\begin{align}
   ds^2
   =
   -
   N^2dt^2
   +
   h_{\alpha\beta}
   \left(dx^{\alpha}+N^{\alpha}dt\right)
   \left(dx^{\beta}+N^{\beta}dt\right)
   .
\end{align}
The extrinsic curvature is
\begin{align}
   K_{\alpha\beta}
   =
   -
   \frac{1}{2N}\left(
      \partial_th_{\alpha\beta}
   -  D_{\alpha}N_{\beta}
   -  D_{\beta}N_{\alpha}
   \right)
   ,
\end{align}
where $D_{\alpha}$ is the extrinsic curvature with respect to the spatial slice.

Puncture initial data is spatially conformally flat and
maximally sliced ($K=0$), and sets $h_{\alpha\beta}=\psi^4\delta_{\alpha\beta}$,
that is the initial spatial metric is conformally flat. 
The extrinsic curvature is specified
by choosing a set of effective black hole masses $m_{(n)}$, 
spins $S^{\gamma}_{(n)}$, momenta $P^{\gamma}_{(n)}$, and locations.
One then solves the Hamiltonian constraint for $\psi$, which then gives us 
$h_{\alpha\beta}$ (the momentum constraint is solved using an analytic formula).
Puncture initial data does not specify the lapse $N$ and shift $N^{\gamma}$.
We set $N^{\alpha}=0$, and choose $N$ to be 
(we set the \texttt{initial-lapse} parameter to \texttt{twopunctures-averaged}
in the \texttt{TwoPunctures} code \cite{Ansorg:2004ds})
\begin{align}
   N
   =
   \left(1 + \frac{m_1}{2r_1} + \frac{m_2}{2r_2}\right)^{-1}
   ,
\end{align}
where $r_i$ is the radial (Euclidean) distance from the $i^{th}$ puncture.
To recover the metric initial data from the ADM variables, we invert
the definitions to get
\begin{align}
   g_{tt}
   =
   -
   N^2
   ,
   \qquad
   &g_{t\alpha}
   =
   0
   ,\nonumber\\
   g_{\alpha\beta}
   =
   h_{\alpha\beta}
   ,\qquad
   &\partial_tg_{tt}
   =
   -
   2N\partial_tN
   ,\\
   \partial_tg_{t\alpha}
   =
   0
   ,\qquad
   &\partial_tg_{\alpha\beta}
   =
   -
   2N K_{\alpha\beta}
   \nonumber.
\end{align}

In puncture coordinates, the black hole apparent horizon is located at $r=m/2$.
We then initially excise an ellipsoid inside that surface 
on our $t=0$ slice after
the \texttt{TwoPunctures} code has solved for the conformal 
factor and interpolated the result on the initial Cartesian grid we use.
The MGH parameters $\hat{g}^{ab}$, $\tilde{g}^{ab}$, and $H_a$ determine 
$\partial_tN$ and $\partial_tN_{\alpha}$.

We set $S^{\gamma}_{(1,2)}=0$, so that the black holes are initially nonspinning.
We choose quasi-circular initial data for the momenta $P^{\gamma}_{(1,2)}$.
In particular, given $r$ and $m_{(n)}$, we set 
(here using spherical polar coordinates) 
\begin{align}
   P^{\gamma}_{(n)}\partial_{\gamma}
   =
   m_{(n)}
   \times
   \left(
      \dot{r}\partial_r
      +
      r \Omega \partial_{\phi}
   \right) .
\end{align}

We choose $\dot{r}$ to be accurate to $2.5$PN order for a quasi-circular
binary, that is it incorporates
the leading-order radiation reaction term, and
we choose $\Omega$ to be accurate to $2$PN order for a
quasi-circular binary \cite{Blanchet:2013haa,Arun:2004ff,vasilis_notebook}.

We note that Kovacs~\cite{Kovacs:2021lgk} 
has recently constructed a more general
set of puncture initial data for black holes in sGB gravity, 
which reduces to the original puncture
data for GR that we use here when one chooses the initial values
of $\phi=\partial_t\phi=0$.

%=============================================================================
\section{\label{sec:perturbative_method}Perturbative solutions
to sGB gravity}
Here, we briefly review the perturbative approach to solving
the equations of motion in shift-symmetric ESGB (sGB) gravity.
While we do not employ the perturbative method in this work
(instead, we solve the full sGB equations of motion),
all previous numerical relativity work comparing to PN theory has 
\cite{Witek:2018dmd,Shiralilou:2020gah,Shiralilou:2021mfl}.
As in those earlier results, we find that at a given frequency,
the amplitude of our scalar waveforms are very similar to the scalar
waveforms produced in the decoupling limit;
however, here we are able  
to directly measure the extra dephasing of the binary black holes
due to the emission of scalar radiation.
This can be traced to the fact that corrections to the scalar
amplitude beyond the leading order decoupling limit scale
as the coupling to the third power, which we show here.

In the perturbative approach, the scalar field and tensor
field are expanded order by order in a small parameter $\epsilon$:
\begin{subequations}
   \label{eq:perturbative_expansion_fields}
   \begin{align}
      g_{ab}
      &=
      \sum_{k=0}^{\infty}\epsilon^kg_{ab}^{(k)}
      ,\\
      \phi
      &=
      \sum_{k=0}^{\infty}\epsilon^k\phi^{(k)}
      .
   \end{align}
\end{subequations}
We assume $\epsilon\sim\lambda/m_1^2=\zeta_1$, and
set $\phi^{(0)}=0$, so that the ``background'' spacetime is vacuum GR.
To zeroeth order in the coupling, the tensor and scalar equations of motion are
\begin{subequations}
   \begin{align}
      G_{ab}^{(0)}
      -
      \nabla_a\phi^{(0)}
      \nabla_b\phi^{(0)}
      +
      \frac{1}{2}g_{ab}^{(0)}\left(\nabla\phi^{(0)}\right)^2
      &=
      0
      ,\\
      \Box^{(0)}\phi^{(0)}
      &=
      0
      ,
   \end{align}
\end{subequations}
where $G_{ab}$ is the Einstein tensor.
We see that if for initial data we set $\phi^{(0)}=\partial_t\phi^{(0)}=0$,
then $\phi^{(0)}=0$ for all time, and the metric field satisfies
the Einstein equations.
From now on we assume $\phi^{(0)}=0$.
To linear order in $\epsilon$, the equations of motion are
\begin{subequations}
   \begin{align}
      G_{ab}^{(1)}
      &=
      0
      ,\\
      \Box^{(0)}\phi^{(1)}
      +
      \lambda\mathcal{G}^{(0)}
      &=
      0
      .
   \end{align}
\end{subequations}
We see that the equation of motion for $g_{ab}^{(1)}$ is also the
vacuum Einstein equations. We can then consistently set $g_{ab}^{(1)}=0$.
The scalar field $\phi^{(1)}$ is no longer zero, even if one initially
sets $\phi^{(1)}=\partial_t\phi^{(1)}=0$ for initial data, as generically
$\mathcal{G}^{(0)}\neq0$.
Solving for $\phi^{(1)}$ to this order, while solving for $g_{ab}^{(0)}$
from the Einstein equations, is called the decoupling approximation
\cite{Witek:2018dmd}.
To second order in $\epsilon$, we have
\begin{subequations}
   \label{eq:scnd_order_pert}
   \begin{align}
      G_{ab}^{(2)}
      -
      \nabla_a\phi^{(1)}
      \nabla_b\phi^{(1)}
      +
      \frac{1}{2}g_{ab}^{(0)}\left(\nabla\phi^{(1)}\right)^2
      +  
      2\lambda
      \delta^{efcd}_{ijg(a}g_{b)d}\left(R^{ij}{}_{ef}\right)^{(0)}
      \nabla^g\nabla_c\phi^{(1)}
      &=
      0
      ,\\
      \Box^{(0)}\phi^{(2)}
      &=
      0
      .
   \end{align}
\end{subequations}
The scalar equation follows from $g_{ab}^{(1)}=0$.
Note that the scalar equation for $\phi^{(2)}$ would have corrections
if the Gauss-Bonnet coupling was nonlinear in $\phi$; for more discussion
see for example Sec II.B.5 in \cite{Witek:2018dmd}.
We see that we can consistently set $\phi^{(2)}=0$.
To third order in perturbation theory, we have
\begin{subequations}
   \begin{align}
      G_{ab}^{(3)}
      &=
      0
      ,\\
      \Box^{(0)}\phi^{(3)}
      +
      \Box^{(2)}\phi^{(1)}
      +
      \lambda\mathcal{G}^{(2)}
      &=
      0
      .
   \end{align}
\end{subequations}
We can set $g_{ab}^{(3)}=0$, but there is a nontrivial correction
to $\phi^{(3)}$ (there would be corrections to $g_{ab}^{(3)}$
if the scalar Gauss-Bonnet coupling was nonlinear in $\phi$,
due to corrections in $\phi^{(2)}$; see the discussion below
Eqs~.\eqref{eq:scnd_order_pert}).
Thus, once one can computed $\phi^{(1)}$, corrections to the scalar
waveform do not appear until $\phi^{(3)}$.
We considered $\zeta_1\sim0.1$ at the largest, so the largest correction
due to nonlinear effects to the amplitude would be of relative order
$\zeta_1^3/\zeta_1=\zeta_1^2\sim0.01$, a $1\%$ effect. This is consistent with what
we see in Figs.~\ref{fig:comparison_sf_pn}
and \ref{fig:rescaling_nonlinearity_phi}.

While nonlinear effects in $\zeta_1$
are not expected to dramatically change the
amplitude of the scalar field during inspiral for sGB gravity, 
nonlinear effects must
be incorporated to determine the long-time dephasing of the binary
due to the emission of scalar radiation. Nonlinear effects
may additionally 
change the spacetime geometry of the merger in ways not captured in
the perturbative approach.
Finally, if the scalar Gauss-Bonnet coupling is not linear in $\phi$,
higher order corrections in the coupling can enter in the scalar
waveform at order $\zeta_1^2$,
and so could be more important in determining the properties of
black hole binaries.

%=============================================================================
\bibliography{../mod_grav}

\end{document}